\documentclass[twocolumn,aps,amssymb]{revtex4}
\usepackage{graphicx}

\begin{document}
\title{{Spin dynamics in La$_{1-x}$Sr$_{x}$MnO$_{3}$ $(x \le 0.175)$
investigated by high-field ESR spectroscopy}}
\author{D. Ivannikov, M. Biberacher, H.-A. Krug von Nidda, A. Pimenov$\footnote{
corresponding author, email:
Andrei.Pimenov@Physik.Uni-Augsburg.DE}$,  and A. Loidl}
\affiliation{Experimentalphysik V, Elektronische Korrelationen und
Magnetismus, Institut f\"{u}r Physik, Universit\"{a}t Augsburg,
86135 Augsburg, Germany} %
\author{A. A. Mukhin}
\affiliation{General Physics Institute of the Russian Acad. of
Sciences, 119991 Moscow, Russia}%
\author{A. M. Balbashov}
 \affiliation{Moscow Power Engineering Institute, 105835 Moscow, Russia}
\date{February 1, 2002}

\begin{abstract}
High-field electron spin resonance (ESR) experiments have been
carried out in single crystals of La$_{1-x}$Sr$_x$MnO$_3$ in the
concentration range $0\le x \le 0.175$. Different quasioptical
arrangements have been utilized in the frequency range
$40<\nu<700$\,GHz and for magnetic fields $B\le12$\,T. A splitting
of the antiferromagnetic resonance (AFMR) mode is observed in the
magnetic field for the parent compound LaMnO$_3$ in agreement with
the antiferromagnetic structure of this material. Abrupt changes
in the AFMR frequencies have been observed around $x\simeq 0.025$
and attributed to the possible transition between
antiferromagnetic and canted state. For increasing Sr-doping the
observed AFMR modes are splitted even in zero field, which can be
naturally explained using a concept of a canted magnetic structure
for $x<0.1$. In La$_{0.825}$Sr$_{0.175}$MnO$_3$ the ESR spectra
are consistent with the ferromagnetic and metallic state. The
lines of ferromagnetic resonance and  ferromagnetic antiresonance
can be clearly seen in the spectra. For intermediate
concentrations $0.1\le x \le 0.15$ a complicated set of ESR
spectra has been observed, which can be well explained by a single
ferromagnetic resonance mode and taking into account
electrodynamic effects.
\end{abstract}

\maketitle

\section{Introduction}

The main physical properties of doped manganites have been
investigated nearly half a century ago by Jonker and van Santen
\cite{jonker} and Wollan and Koehler \cite{wollan}. The observed
doping and temperature dependences have been qualitatively
explained by de Gennes \cite{degennes} using an interplay of
Mn-O-Mn superexchange and Zener's double exchange (DE) mechanism
\cite{zener}. In this model, on increasing doping the
antiferromagnetic (AFM) and insulating (I) LaMnO$_{3}$ passes
through a canted AFM (CAF) ground state and arrives at a
ferromagnetic (FM) and metallic (M) state at doping levels
$x\gtrsim 0.2$. Within this picture a phase diagram of manganites
has been constructed, which was in agreement with the existing
experimental data.

The interest in doped manganites has been considerably renewed
after several groups had observed exceptionally high values of the
magnetoresistance in thin films \cite{gmr}, later termed as
"colossal magnetoresistance" (CMR). The CMR effects at the
ferromagnetic phase transition were analyzed within an extended
double-exchange model \cite{furukawa} but also with models taking
a strong electron-phonon coupling \cite{millis95,roder} or a
percolative metal-insulator transition \cite{moreo99} into
account. Subsequently it was realized that the rich phase diagram
observed in various manganite systems can only be explained
considering additional degrees of freedom such as Jahn-Teller
distortions \cite{millis95,roder}, electronic correlations and
charge and orbital order \cite{maezono,kilian,horsch,brink}.

In this paper we present high-field (HF) electron spin resonance
(ESR) investigations of single crystals of La$_{1-x}$Sr$_x$MnO$_3$
for concentrations  $0\le x \le 0.175$, ranging from the AFM
insulator to the FM metallic regime. The paper is organized as
follows. The next section discusses the problem of phase
separation in manganites. The discussion is focused on the
low-doping region of the phase diagram, where important
information can be extracted from the high-field ESR-experiments
\cite{mukhin00,pimenov00}. The subsequent section presents the
details of the sample preparation and the experimental procedure.
The experimental results are discussed and presented within three
groups, $0\le x<0.1$, $0.1\le x \le 0.15$, and $x>0.15$,
respectively. The first region can be well described as
antiferromagnetic insulator with a (modulated) canted structure
followed by a ferromagnetic insulating ground state and, finally,
by the ferromagnetic metal for $x>0.15$.

\section{\label{seccanted}Phase Separation vs. Canted Structure}

Enormous theoretical and experimental efforts have been devoted to
the problem of phase separation in manganites. Already Wollan and
K\"{o}hler \cite{wollan} have observed the coexistence of ferro-
and antiferromagnetic lines in the neutron-scattering spectra, an
effect, which could be equally explained by the canting of the
parent antiferromagnetic structure or by phase separation into
ferro- and antiferromagnetic domains. Later on, de Gennes derived
theoretically \cite{degennes} the canted antiferromagnetic
structure as a ground state of low-doped manganites. A number of
subsequent experimental results have since then been explained
using the concept of a canted structure
\cite{jirak,kawano,maiti,hennion97,okimoto,brion,skumryev}.

However, in recent years a number of theoretical models predicted
that the CAF structure becomes unstable against electronic phase
separation into ferromagnetic (FM) and antiferromagnetic (AFM)
regions \cite{moreo99,nagaev,nagaev98,yunoki}. According to the
calculations on the basis of different models, the phase
separation is energetically favorable in the whole range of the
phase diagram. At present a number of reviews devoted especially
to this problem is available in the literature
\cite{dagotto01,kagan01}.

Phase diagrams of doped manganites are very complex and in many
cases difficult to interpret. It is therefore reasonable to expect
that different physical mechanisms play a major role at different
parts of the phase diagram and the answer to the PS-CAF question
will be ambiguous. Postponing for a moment the discussion of the
experimental data, the situation is not simple even from the
theoretical point of view. For example, analyzing lattice effects
on the magnetism of LaMnO$_3$ using the local-spin-density
approximation, Solovyev \textit{et al.} \cite{solovyev96} find a
stable canted phase. Investigating the phase diagram of manganites
on the basis of the DE model, Nagaev \cite{nagaev98} confirmed the
conclusions of de Gennes \cite{degennes}: the CAF structure can be
energetically favorable for certain portions of the phase diagram.
Parameter regions resulting in a stable CAF phase have also been
found using a Schwinger-boson representation for the magnetic
moments \cite{arovas}. More recently, similar results have been
obtained re-investigating the DE theory within either
multiple-scattering theory \cite{solovyev01} or within a model
including Coulomb repulsion and electron-phonon coupling
\cite{zou}. However, we recall that the stability of a CAF phase
in manganites has been obtained by comparative minority of
theoretical concepts dedicated to the topic of phase separation.

A large body of experimental work has been devoted to the
detection of the phase separation in manganites, which led in many
cases to convincing experimental evidences \cite{mori,radaelli}
especially in the vicinity of the ferromagnetic phase transition
\cite{uehara,fath}. The discussion of most of these results has
been recently reviewed by Dagotto \textit{et al.}
\cite{dagotto01}. In spite of enormous experimental progress in
the field of phase separation, the
 region of low-doped LaMnO$_3$, and especially the composition range
$x<0.1$ still remains to be the matter of discussion. Among
others, the Sr- and Ca-doped LaMnO$_3$ represent the most
extensively studied systems. Their physical properties are quite
similar, especially in the low-doping regime. One of the most
prominent differences is a slightly stronger effect of
Sr-substitution: the transition to a ferromagnetic insulator takes
place around $x\simeq 0.1$ for Sr \cite{parask00,liu}, compared to
$x\simeq 0.125$ for Ca-substitution \cite{hennion00,hennion01}.

First neutron scattering data measured on Ca-doped LaMnO$_3$ by
Wollan and Koehler \cite{wollan} have been interpreted as equally
compatible with either PS or CAF picture in the low-doped region.
About forty years later, after the interest to manganites has been
renewed, Kawano \textit{et al.} \cite{kawano} explained the
neutron-scattering and magnetization results as pointing towards
the CAF magnetic structure. Later on, using the same technique,
Hennion \textit{et al.} \cite{hennion97,hennion98} found evidences
for inhomogeneities in the canted structure, which have been
termed "magnetic droplets" or "magnetic polarons". It should be
pointed out that these results have been interpreted using the CAF
phase as a ground state with the droplets arising from the
effective modulation of the canting angle. This picture of the
modulated CAF phase has been  later confirmed by the same group
\cite{hennion00,hennion01}. Almost simultaneously with the
neutron-scattering data, evidences for inhomogeneities in the
low-doped region of the phase diagram have been found by NMR
spectroscopy by Allodi \textit{et al.} \cite{allodi97,allodi98}.
The NMR spectra pointed towards pure phase segregation into FM and
AFM domains (see also Refs. \cite{kumagai,savosta}). The NMR
results seemingly contradict other available experimental data on
low-doped manganites, which point towards the CAF structure as a
ground state. However, as discussed by Kumagai \textit{et al.}
\cite{kumagai}, the NMR spectra might be affected by the ceramic
character of the samples. This open problem certainly deserves
additional investigations.

The antiferromagnetic-resonance (AFMR) technique has been applied
to Sr-doped LaMnO$_3$ in the low-doping regime $(0\le x \le 0.1)$
without magnetic field by Mukhin \textit{et al.} \cite{mukhin00}.
In addition, these investigations have been carried out in
magnetic fields on untwinned single crystals of
La$_{1-x}$Sr$_x$MnO$_3$ with $x=0.05$ \cite{pimenov00}. Two
resonance modes have been observed in the spectra, which revealed
distinct excitation conditions, magnetic field and doping
dependence. These results could be directly explained using the
CAF picture and have been found to contradict the phase separation
into pure FM and AFM phases. In agreement with this conclusion,
the recent calculations of the spin-wave branches by Rom\'{a}n and
Soto \cite{roman} confirm the distinctive character of the AFMR
experiments on solving the PS-CAF problem. However, the AFMR
experiments do not exclude a possible modulation of the canting
angle, as found in neutron-scattering experiments
\cite{hennion00,hennion01}. The modulation rather produces an
additional broadening of the observed lines
\cite{mukhin00,pimenov00}.

The problem of phase separation has been addressed from the point
of view of magnetization data by Paraskevopoulos \textit{et al.}
\cite{parask00} and Geck \textit{et al.} \cite{geck}. Although the
magnetization measurements cannot in general case distinguish
between PS and CAF structures, the detailed analysis of the
anisotropic magnetization \cite{parask00} or hysteresis curves
\cite{geck} did provide weighty arguments in  favor of a CAF
structure.

Concluding the discussion on the PS-CAF problem for low-doped
manganites, we try to formulate the description of the magnetic
structure for the low-doped manganites, which seems to be
consistent with the majority of existing experimental data: The
magnetic ground state is represented by a canted magnetic
structure, which is statically modulated around an average value.
Therefore, the term \emph{modulated canted structure} seems to be
appropriate in this case. Such idea of the local inhomogeneities
of spin distortion has already been discussed by de Gennes
\cite{degennes}. Naively, in weakly doped manganites every doping
atom could be considered as a center, which tends to locally
increase the canting angle and thus to produce the modulation of
the CAF structure. However, according to the neutron scattering
data \cite{hennion98,hennion00}, the inhomogeneities in the
magnetic structure are roughly one order of magnitude larger than
the distance between the doping ions.

\section{Experimental Details}

Single crystals of La$_{1-x}$Sr$_{x}$MnO$_{3}$ were grown by the
floating-zone method with radiation heating \cite{preparation}.
Raw La$_2$O$_3$, SrCO$_3$ and Mn$_3$O$_4$ chemicals of a high
purity (not less 99.99\%) were used for ceramic rods preparation.
Some excess of Mn$_3$O$_4$ concentration ($\sim 0.5$ at.\%) was
used in order to compensate a Mn loss due to the evaporation from
the melt in the floating-zone process. The initial synthesis of
composition was provided by annealing of mixed chemical powder at
a temperature about $1200^o$\,C during 24 h. After pressing of
feed rods they were sintered at $1350^o$\,C during 24 h. Single
crystals with concentrations $x\le 0.075$ were grown in Ar
atmosphere, while for $x>0.075$ the samples were grown in air. The
typical growth direction was [110]. In order to obtain crack free
crystals they were annealed at a temperatures about $1300^o$\,C.
X-ray powder-diffraction measurements revealed  single phase
material. Two-dimensional X-ray topography of the crystals
indicated a twin structure of all crystals, except for $x=0.05$,
which turned out to be twin-free.

The temperature dependence of the dc resistivity of
La$_{1-x}$Sr$_x$MnO$_3$ \cite{mukhin98} was measured utilizing a
four-point technique and resembles quantitatively and
qualitatively the data known from literature
\cite{moritomo,urushibara}. In addition, the submillimeter-wave
properties \cite{pimenov99} and magnetization
\cite{paraskmmm,parask00} of these samples have been investigated
and have been published previously. Plane-parallel plates of
approximately 8$\times $8$\times 1$\,mm$^{3}$ have been cut from
the crystals for the high-field ESR measurements.

The high-field ESR spectra for frequencies 40 GHz $\leq \nu \leq $
700 GHz ($1.3-23$ cm$^{-1}$) were recorded using a quasioptical
technique utilizing backward-wave oscillators as coherent light
sources \cite{volkov}. Depending upon the transparency of the
samples, two different types of experiments have been employed,
i.e. transmission and reflection geometry. Using a similar
technique, high-field ESR spectra of Nd$_{1-x}$Ca$_x$MnO$_3$ have
been recently investigated by Dupont \textit{et al.}
\cite{dupont}.

Within the transmission geometry the conventional quasioptical
arrangement \cite{volkov} is equipped with a superconducting
split-coil magnet.  The Mylar optical windows allow to carry out
transmission experiments in magnetic fields up to 8\,T with the
field parallel (Faraday geometry) as well as perpendicular (Voigt
geometry) to the propagation of the electromagnetic beam
\cite{zvezdin}. The combination of both geometries is especially
important investigating the ferromagnetically-ordered samples
because in both cases different resonance conditions occur due to
demagnetization fields. For a thin platelet the following
expression for the ferromagnetic resonance frequencies can be
written:
\begin{eqnarray}
\omega _{FM}=\gamma \sqrt{(H+4\pi M)H}  \ \, \ \textrm{Voigt
geometry} \nonumber
\\
( H \parallel \textrm{to the platelet surface})\label{voigt},
\\
\omega _{FM}=\gamma \left| H-4\pi M\right|  \ \, \ \textrm{Faraday
geomerty} \nonumber
\\
( H \perp \textrm{to the platelet surface})\label{faraday} ,
\end{eqnarray}
where $\gamma$ is the gyromagnetic ratio and $M$ is the
magnetization. Equations (\ref{voigt}) and (\ref{faraday}) have
been written under the assumption that the platelet is oriented
perpendicular to the quasioptical beam (i.e. normal incidence).
Therefore, the resonance frequencies are shifted  to higher
frequencies for the Voigt  and to  lower frequencies for the
Faraday geometry.

The simplicity of the transmission geometry allows two different
experimental modes: frequency-sweep and field-sweep runs. In the
frequency-sweep mode the data can be collected even without
magnetic field, which strongly simplifies the detection of weakly
field-dependent modes. The frequency-dependent transmission
spectra are analyzed using the Fresnel formulae for the
transmittance $T=|t|^2$ of a plane-parallel sample (Voigt
geometry) \cite{born}
\begin{eqnarray}
\label{eqtran}  \lefteqn{\hspace{2cm} t=\frac{(1-r^{2})t_{1}}{1-r^{2}t_{1}^{2}} \ , }\\
& & \text{where\ }r=\frac{\sqrt{\varepsilon /\mu
}-1}{\sqrt{\varepsilon /\mu }+1}\ \text{ and } t_{1}=\exp (-2\pi
i\sqrt{\varepsilon \mu }\,d/\lambda) \ \text{.}\nonumber
\end{eqnarray}
Here $r$ is is the reflection amplitude at the air-sample
interface, $t_1$ is the ``pure''  transmission amplitude,
$\varepsilon$ and $\mu$ are the (complex) dielectric permittivity
and magnetic permeability
 of the sample, respectively, $d$ is the sample
thickness, and $\lambda$ is the radiation wavelength. This
expression can be applied also for anisotropic crystals, if  the
incident radiation is polarized along the principal optical axes.
For experiments in Faraday geometry the transmission has to be
determined by a more complicated expressions, which strongly
depends
 on the mutual orientation of analyzer and polarizer and
will be presented in a separate paper.

The relatively good transparency of the low-doped ($x<0.1$)
manganites in the submillimeter frequency range resulted in
interference patterns due to internal reflections from the sample
surface. The observation of these interferences allowed the
calculation of the optical parameters of the sample without
measuring the phase shift of the transmitted signal. The
dispersion of the magnetic permeability has been taken into
account assuming a harmonic oscillator model for the complex
magnetic permeability:
\begin{equation}
\mu ^{*}(\nu )=\mu _{1}+i\mu _{2}=1+\frac{\Delta \mu \, \nu
_{0}^{2}}{\nu _{0}^{2}-\nu ^{2}+i\nu \delta}  \label{eqreson}
\end{equation}
where $\nu _{0}$, $\Delta \mu $ and $\delta$ are eigenfrequency,
mode strength and width of the resonance, respectively. The
dielectric parameters of the sample
$(\varepsilon^{*}=\varepsilon_1+i\varepsilon_2)$  in good
approximation behave regularly in the vicinity of the resonance
frequency. Hence, frequency-sweep measurements allowed to obtain
absolute values of the parameters of the ESR-AFMR lines.

The reflection geometry has been utilized in a top-loading 16\,T
superconducting magnet. The usual quasioptical elements have been
set-up outside the magnet. A stainless-steel rod has been
constructed to guide the beam inside the magnet, which was finally
focused on the sample by a Teflon lens. This arrangement did not
allow to change the sample against the reference mirror, which
hampered frequency-sweep experiments. In the reflection experiment
with the top-loading magnet the sample surface is perpendicular to
the static magnetic field and to the propagation direction of the
beam, which corresponds to the Faraday geometry.

\section{Results and Discussion}

Since the early phase diagrams \cite{bogush,urushibara} of
low-doped manganites a number of additional details and
corrections have been provided \cite{kawano,yamada,zhou}. On the
basis of the samples investigated in this work, a similar phase
diagram has been presented, which has been constructed on the
basis of conductivity and magnetization measurements
\cite{mukhin98,parask00}. Except for minor details and slightly
different interpretations, this diagram agrees well with the
established ones and also with the detailed phase diagram,
published recently by Lui \textit{et al.} \cite{liu}. These phase
diagrams will be referred when discussing the results of the
high-field ESR experiments.

\subsection{The Parent Antiferromagnet ($x=0$)}

\begin{figure}[]
\centering
\includegraphics[width=6cm,clip]{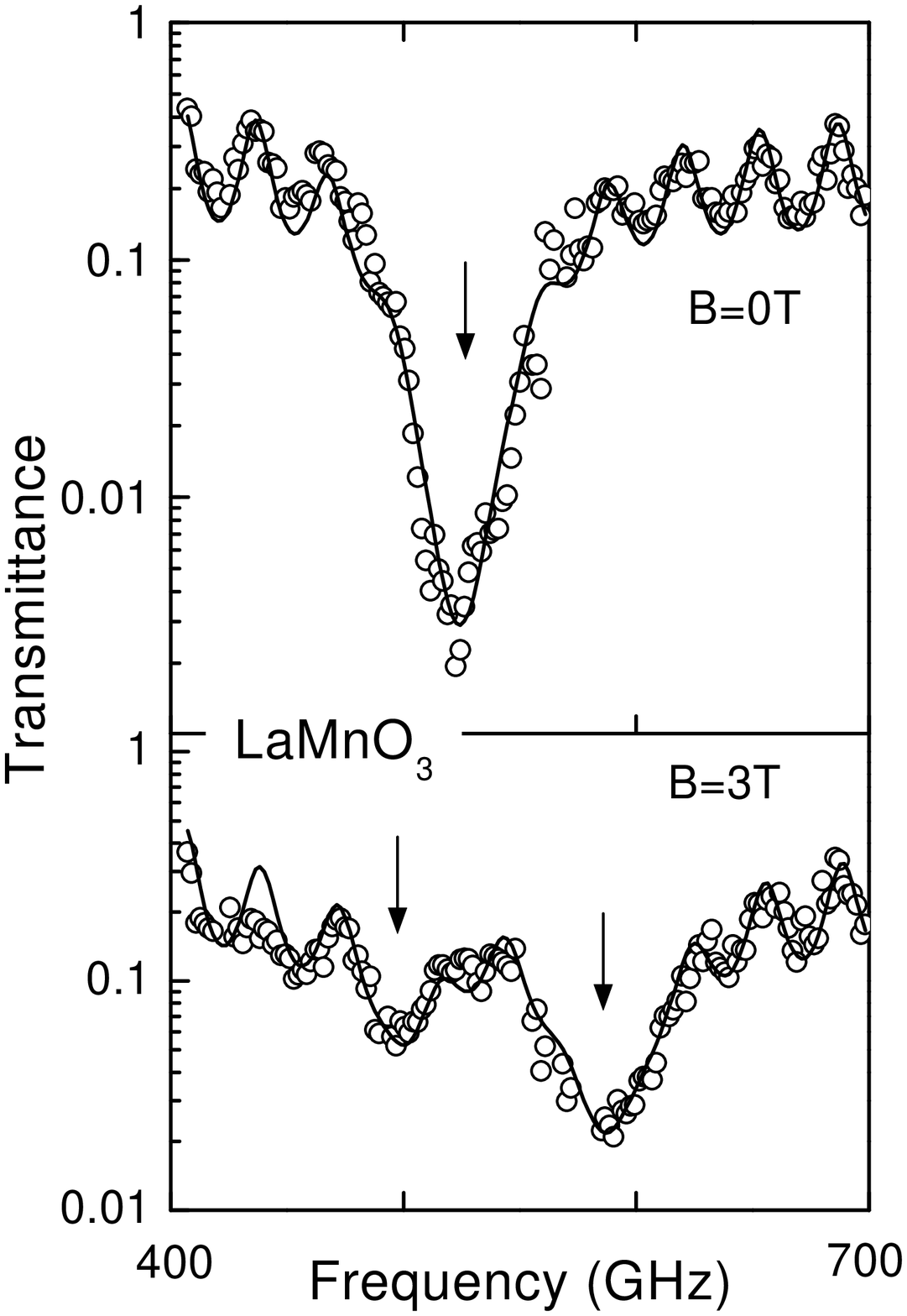}
\vspace{0.2cm} \caption{Antiferromagnetic resonance observed in LaMnO$_{3}$ without
magnetic field (upper panel)
 and in a static field with $B=3$\,T. Symbols - experiment,
lines - calculations on the basis of the Fresnel equations.} \label{fig00f}
\end{figure}

Fig.\ \ref{fig00f} shows the frequency dependence of the
transmittance of LaMnO$_3$ in the frequency range of the
antiferromagnetic resonance. The spectra in zero magnetic field
are dominated by an intensive absorption line around 520\,GHz,
which represents the AFMR-mode. The transmittance spectra can be
well approximated using a Lorentzian (Eq.\ \ref{eqreson}) for the
resonance mode. The fringes on  both sides of the resonance line
arise due to interference of the beam reflected from the opposite
sides of the sample. The amplitude and the period of these fringes
allow an independent determination of the dielectric parameters of
the sample. The solid lines in Fig.\ \ref{fig00f} represent the
calculated transmittance of a plane-parallel sample according to
the Fresnel equations, Eq.\ (\ref{eqtran}).

A better description of the resonance minimum can be achieved
assuming two closely separated AFMR lines instead of a single line
\cite{mukhin00}. Previously, the small splitting of the AFMR line
has been explained by the magnetic anisotropy due to the low
(orthorhombic) symmetry of the crystal \cite{mukhin00}. However,
it is not clear at present, whether this effect is simply due to a
slight oxygen nonstoichiometry. E.g. the existence of a small
canting has been observed even in pure stoichiometric LaMnO$_3$
\cite{matsumoto} and attributed later \cite{topfer} to a
Dzyaloshinski-Moriya exchange coupling.

\begin{figure}[]
\centering
\includegraphics[width=6cm,clip]{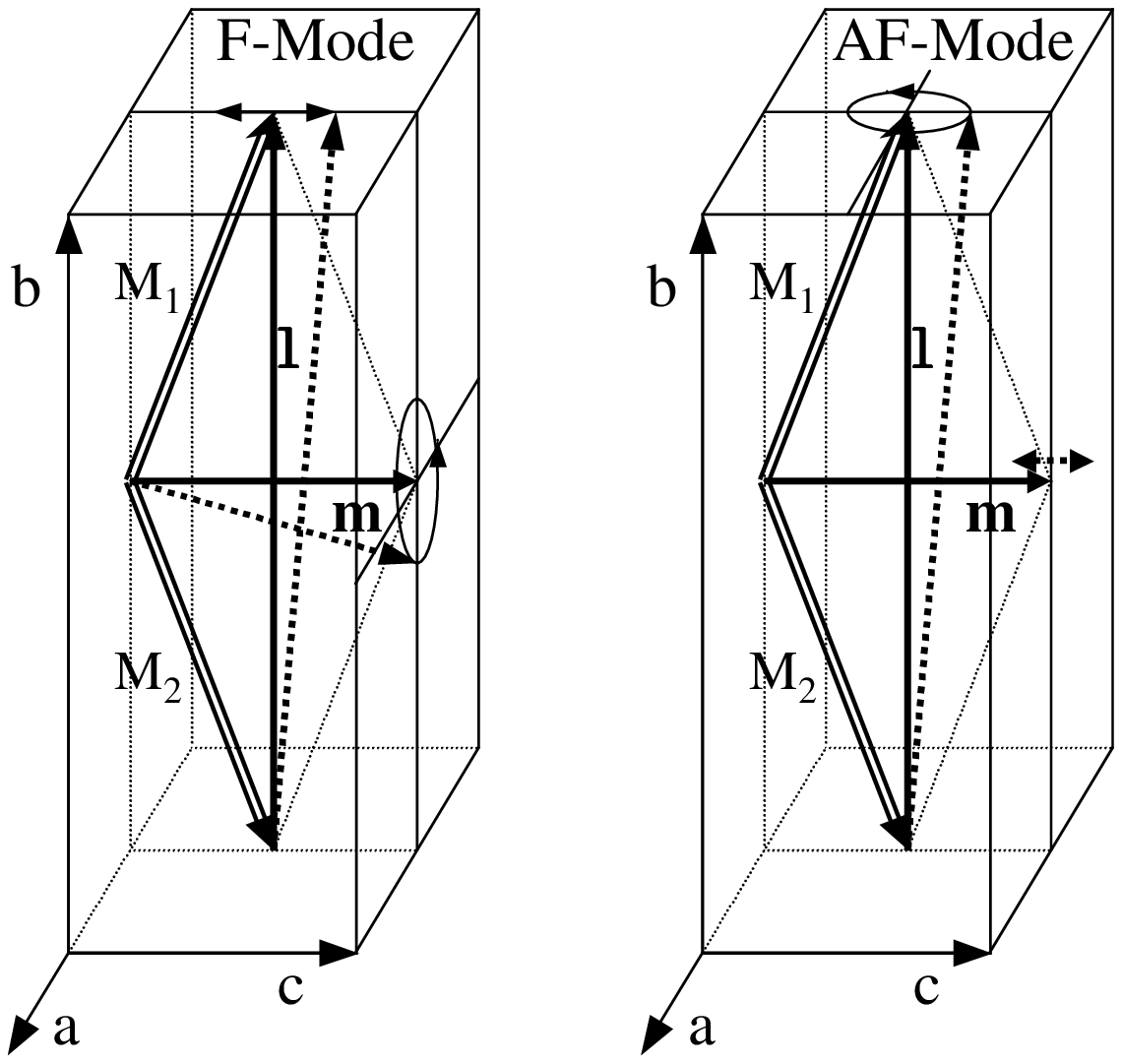}
\vspace{0.2cm} \caption{Schematic representation of the antiferromagnetic modes for a
canted antiferromagnetic structure. The magnetic moments of the two sublattices ${\bf
M}_1$ and ${\bf M}_2$ correspond to adjacent (ab)-layers and are brought to the same
point for simplicity. $\textbf{m}={\bf M}_1 + {\bf M}_2$ - ferromagnetic moment,
$\textbf{l}={\bf M}_1 - {\bf M}_2$ - antiferromagnetic moment. The oscillations
corresponding to two possible modes are also shown (ellipses and double arrows). The
quasi-ferromagnetic mode (F) is excited by $h_{ac}\parallel a$ and $h_{ac}\parallel b$,
the  quasi-antiferromagnetic mode (AF) is excited by $h_{ac}\parallel c$.}\label{figsh}
\end{figure}

The application of the static magnetic field splits the AFMR mode
into two clearly resolved modes (lower frame of Fig.\
\ref{fig00f}). This behavior is typical for an antiferromagnet
\cite{foner,turov} and has previously been observed in field-sweep
spectra in pure LaMnO$_3$ by Mitsudo \textit{et al.}
\cite{mitsido}. Within a simplified picture the line splitting
takes place for the orientation of the external magnetic field
parallel to the magnetic moments of the sublattices (i.e. parallel
to the antiferromagnetic vector, ${\bf l}={\bf M}_1-{\bf M}_2$ in
Fig.\ \ref{figsh}). For a two-sublattice axial antiferromagnet the
external magnetic field along the easy axis removes the degeneracy
of two circularly-polarized modes and splits them into two lines.
The corresponding resonance frequencies $\omega_{\pm}$ are given
in a crude approximation simply by \cite{foner}:
\begin{equation}
\omega_{\pm}=\omega_0 \pm \gamma H \quad.
\end{equation}
Here $\omega_0$ is the resonance frequency without external field,
$H$ is the magnetic field parallel to the antiferromagnetic vector
and $\gamma$  is the gyromagnetic ratio. A closer analysis of the
field-dependent antiferromagnetic resonances in Fig.\ \ref{fig00f}
reveals the appearance of an additional splitting of the lines,
which is due to twinning of the crystal and is documented in Fig.\
\ref{figd} (Section \ref{secsum}).

\subsection{\label{seccan} The Canted Antiferromagnet ($0<x<0.1$)}

As discussed in Section \ref{seccanted}, as far as low-doped
manganites are concerned, the concept of a (modulated) canted
magnetic structure is well in agreement with the majority of
experimental data in the field
\cite{wollan,kawano,jirak,mukhin00,pimenov00,hennion97,parask00,liu,
maiti,okimoto,brion,skumryev,paraskmmm,hennion00,hennion01,hennion98,geck,urushibara}.
Specifically for the high-field ESR experiments, this concept
successfully describes the splitting of the antiferromagnetic
lines, excitation conditions, magnetic field, doping dependence,
etc. \cite{mukhin00,pimenov00}. To our best knowledge there exists
no theory, which could explain the AFMR data on the basis of phase
separation into pure FM and AFM regions. Therefore, the following
presentation and discussion will be given within the concept of
the canted magnetic structure, only.

The substitution of the trivalent La$^{3+}$ in the parent
LaMnO$_3$ by divalent Sr$^{2+}$ or Ca$^{2+}$ introduces holes in
the structure. Assuming Zener's double-exchange mechanism
\cite{degennes,zener}, the holes favor the ferromagnetic
orientation of the magnetic lattice and therefore lead to a
canting of the magnetic moments along the crystallographic c-axis.

Compared to the pure antiferromagnetic structure, the modes of the
antiferromagnetic resonance are splitted even in the absence of an
external magnetic field. The resonances of the canted structure
can be well described by separating two sublattices of both
magnetization directions (${\bf M}_1$ and ${\bf M}_2$,
two-sublattice model \cite{moria}). The behavior of the
antiferromagnetic resonances of the canted structure has been
calculated by a number of authors \cite{moria,herrmann,cinader},
and especially for manganites by de Gennes \cite{degennes} nearly
forty years ago.

The two modes observed can be represented as the oscillation of
the ferro- and antiferromagnetic vectors, ${\bf m}={\bf
M}_{1}+{\bf M}_{2}$, ${\bf l}={\bf M}_{1}-{\bf M}_{2}$,
respectively. Here ${\bf M}_{1}$ and ${\bf M}_{2}$ are the
magnetizations of the sublattices. These vectors and the
corresponding modes are represented in Fig. \ref{figsh}. The two
AFMR-modes can be termed quasiferro- (F-mode) and
quasi-antiferromagnetic (AF-mode) resonances, respectively. The
interaction of the modes with the electromagnetic field is
realized vie the term (${\bf m}\cdot H$) in the free energy. The
oscillations in the F-mode involve the following components of the
magnetic vectors: $\textbf{m}_a, \textbf{m}_b$, and $\textbf{l}_c$
(Fig. \ref{figsh}, left panel). This mode can therefore be excited
by the electromagnetic wave with the ac-magnetic field
($\tilde{h}$) having a nonzero component in the ab-plane. By
analogy, the AF-mode (Fig. \ref{figsh}, right panel), which
involves ($\textbf{m}_c, \textbf{l}_a, \textbf{l}_b$) is excited
for ($\tilde{h}$) lying along the c-axis \cite{herrmann,mukhin}.

\begin{figure}[]
\centering
\includegraphics[width=6cm,clip]{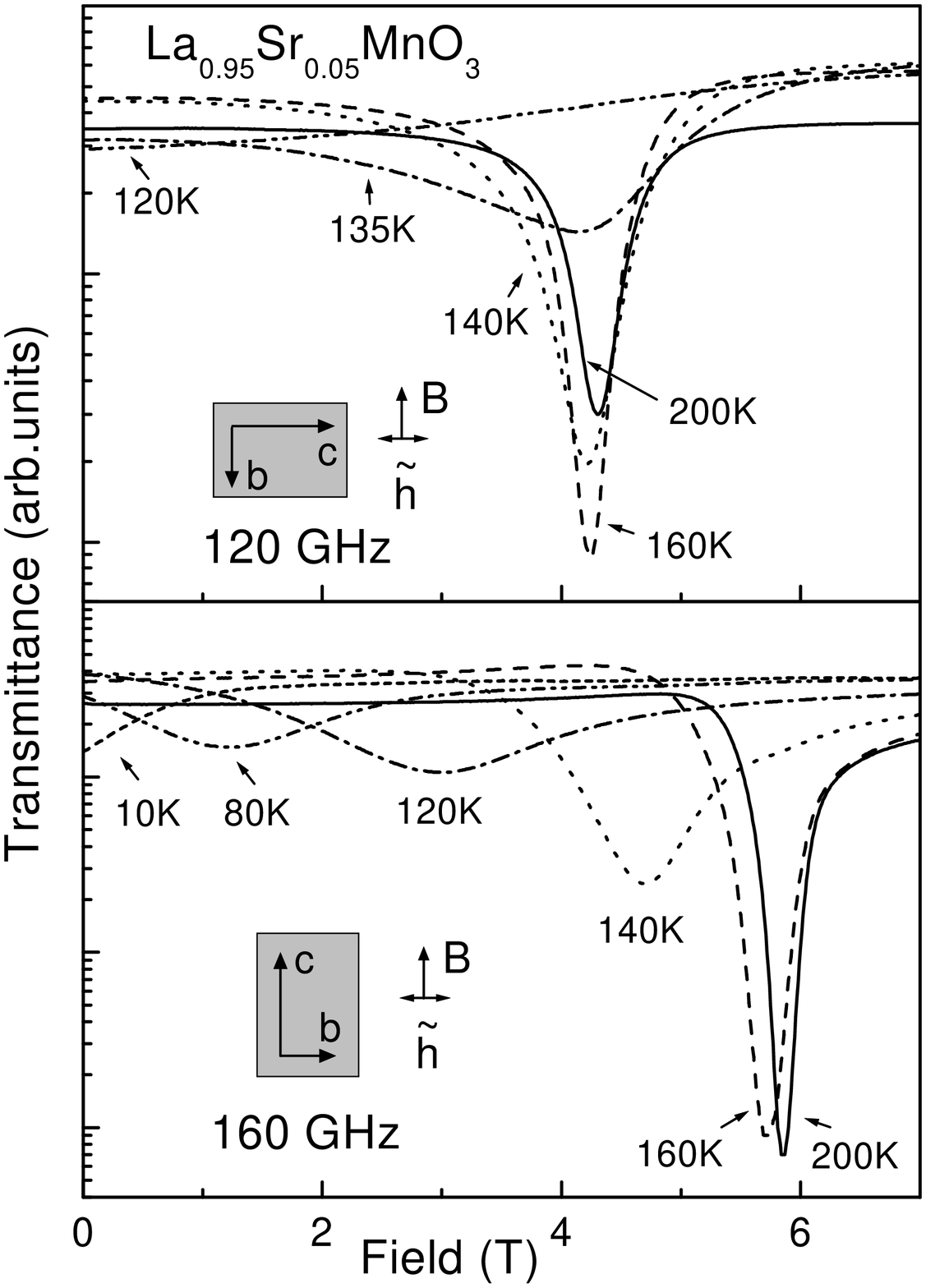}
\vspace{0.2cm} \caption{High-field ESR spectra of an untwinned
La$_{0.95}$Sr$_{0.05}$MnO$_{3}$  for two experimental geometries, as indicated, and for
temperatures below and above $T_{\rm N}\simeq 140$\,K.} \label{fig05m}
\end{figure}

A realistic two-sublattice model for manganites  \cite{mukhin} has
been successfully applied to the doping dependence of the
AFMR-modes in La$_{1-x}$Sr$_x$MnO$_3$ \cite{mukhin00}, and to the
magnetic-field dependence of the resonances in untwinned
La$_{0.95}$Sr$_{0.05}$MnO$_3$ single crystal \cite{pimenov00}. An
example of the transmittance spectra in the field-sweep modus for
 $x=0.05$ is given in Fig.\ \ref{fig05m}, which
represents the field dependence of the submillimeter transmittance
 for two different orientations. Above the antiferromagnetic transition $T_{\rm
N}\simeq 140$\,K a single paramagnetic line (EPR) at $g=2$ is
observed. The width of the EPR line for $\tilde{h} ||$\,c
($\overline{B}\parallel b$, upper panel) is substantially larger
than the line-width for $\tilde{h} ||$\,b ($\overline{B}\parallel
c$, lower panel). Below the magnetic ordering transition the mode
with $\tilde{h} ||$\,b (lower panel of Fig.\ \ref{fig05m}) rapidly
shifts to lower fields and finally stabilizes at $\sim 180$\,GHz
at low temperature and in zero magnetic field. According to the
excitation conditions and the zero-field resonance frequency, this
mode corresponds to the quasi-ferromagnetic mode in Fig.\
\ref{figsh}.

The resonance line in the upper panel of Fig.\ \ref{fig05m}
apparently disappears below the magnetic transition. However, the
analysis of the high-frequency spectra shows that this line
strongly broadens in the field-sweep scale. Similar to the quasi-F
mode, this line shifts  to lower fields and saturates at $\nu \sim
420$\, GHz (for $B=0$\,T). The large difference between this value
and the frequency of the field-sweep experiment (120\,GHz) is the
second reason for the apparent disappearance of the ESR line as
documented in the upper panel of Fig.\ \ref{fig05m}.

\begin{figure}[]
\centering
\includegraphics[width=6cm,clip]{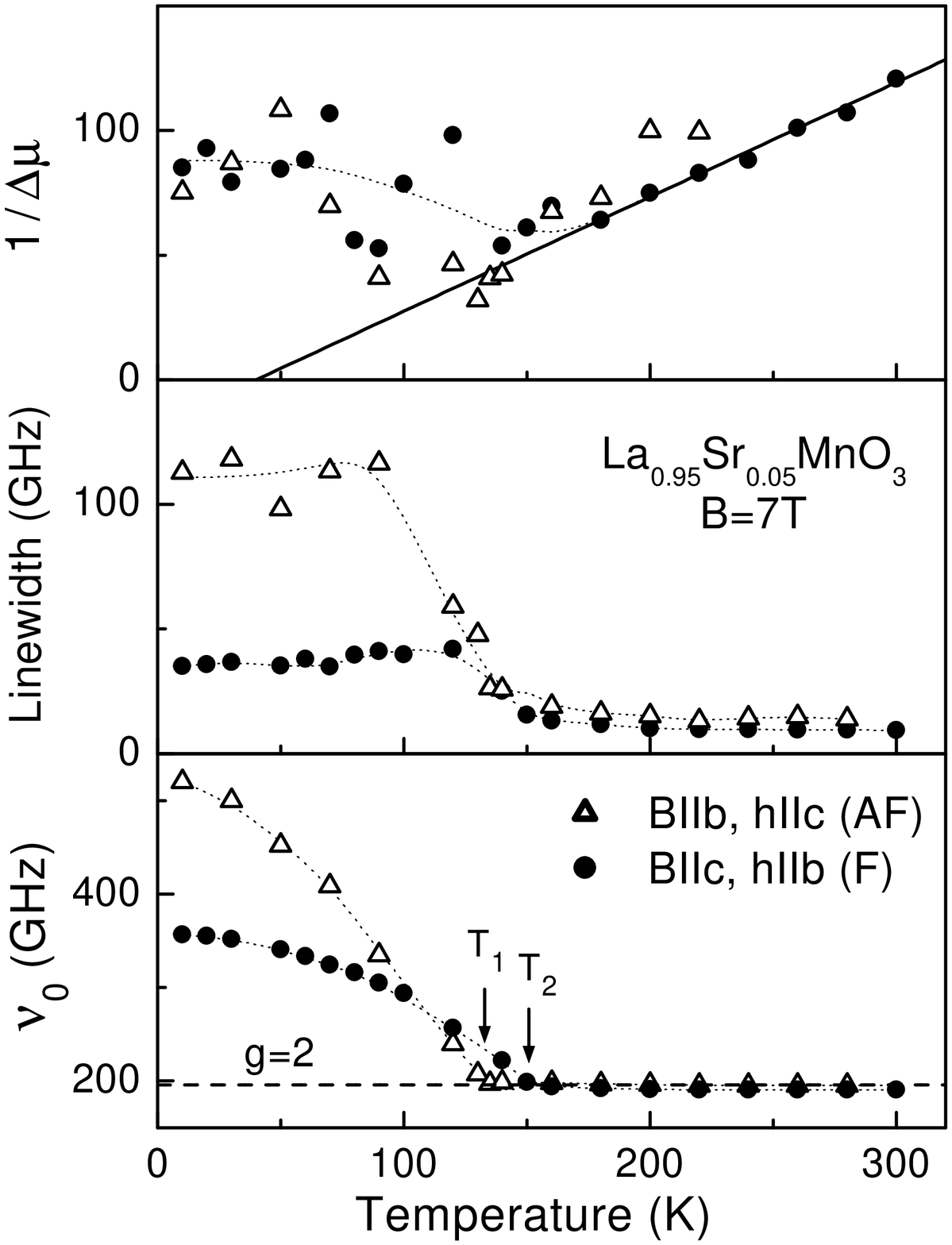}
\vspace{0.2cm} \caption{Temperature dependence of the parameters of the ESR modes in a
La$_{0.95}$Sr$_{0.05}$MnO$_{3}$. The parameters have been obtained analyzing the
frequency dependent transmittance as described in the text. The symbol notation is the
same in all frames. The experimental geometry corresponds to that
of Fig.\ \ref{fig05m}.  Dotted lines are guides to the eye.\\
Upper panel: Inverse line intensity. The solid line for $B||c$ corresponds to a
Curie-Weiss law with $\mu_{eff}\simeq 7\,
\mu_{\rm B}$. \\
Middle panel: Line width.\\
Lower panel: Resonance frequency. The dashed line indicates $g=2$. Arrows mark different
Neel temperatures for the two measuring geometries.} \label{fig05t}
\end{figure}

Fig.\ \ref{fig05t} shows the temperature dependence of the
parameters of the ESR modes of Fig.\ \ref{fig05m}. These data have
been obtained mainly from the analysis of the frequency-dependent
transmittance via Eqs.\ (\ref{eqtran}) and (\ref{eqreson}). As
demonstrated in the lower panel of Fig.\ \ref{fig05t}, in the
paramagnetic state the ESR line is located around $g\simeq2$
independent of the orientation of the magnetic field.  In the
magnetically ordered state the excitation conditions become
orientation-selective (Fig.\ \ref{figsh}) and both geometries
reveal strongly different resonance frequencies. In addition, the
lower panel of Fig.\ \ref{fig05t} reveals that the magnetic-field
induced shift of the ordering temperature is orientation
dependent, too. The corresponding transition temperatures are
marked by $T_{\rm N}=T_1$ and $T_2$, respectively. The magnetic
field along the c-axis favors the low-temperature magnetic
configuration ${\bf m}\parallel c$, ${\bf l}\parallel b$ and
therefore strongly enhances $T_{\rm N}=T_2$. On the contrary, for
$B||b$ the application of the static field favors the
configuration ${\bf m}\parallel b$, ${\bf l}\parallel c$, which
has a lower Neel temperature and is energetically unstable. In
that case the transition at $T_{\rm N}=T_1$ broadens and remains
approximately field-independent.

As is documented in the middle panel of Fig.\ \ref{fig05t}, the
ESR line strongly broadens below $T_{\rm N}\simeq140$\,K. This
explains the non-observability of these lines in conventional
X-band ESR \cite{ivanshin} and the apparent disappearance of the
$B||b$ mode in the upper panel of Fig.\ \ref{fig05m}. The
anisotropy of the line-width at low temperatures is seen both in
the paramagnetic and in the magnetically ordered state. From an
extensive analysis of X-band (9 GHz) ESR experiments
\cite{ivanshin,deisenhofer}, this anisotropy has been attributed
to the Dzyaloshinski-Moriya interaction and to the crystal-field
effects. A detailed report on the low-frequency EPR-experiments in
La$_{1-x}$Sr$_x$MnO$_3$ has been published elsewhere
\cite{deisenhofer,ivanshin}.

The upper panel of Fig.\ \ref{fig05t} shows the mode contribution
of the ESR modes in La$_{0.95}$Sr$_{0.05}$MnO$_3$. At low
temperatures the contributions of both modes coincide within the
experimental accuracy. The absolute values of the contribution
($\Delta \mu \sim 0.01$) agree  well with the predictions of the
two-sublattice model \cite{pimenov00}. The mode contribution for
$B||c$ in the paramagnetic state follows the Curie-Weiss behavior.
However, the estimate of the effective paramagnetic moment yields
$\mu_{eff}\simeq 7\,\mu_{\rm B}$, which is substantially higher
than $\mu_{eff}\simeq 5.5\,\mu_{\rm B}$, obtained from
dc-susceptibility \cite{parask00}.
 This effect most probably is
due to the smearing of the Curie-Weiss law in the vicinity of the
Neel temperature and in high magnetic fields ($B=7$\,T).

%An interesting and unexpected behavior is revealed by the AF-mode
%($B||b$) for $T>220$\,K\,: the intensity of this line abruptly
%decreases with increasing temperature above $T=200$\,K. The ESR
%signal has fully disappeared from the spectra already at room
%temperature. We note that similar effect has been observed in the
%X-band (9 GHz) ESR-experiments \cite{ivanshin} and attributed to
%the dominance of the skin-effect in the sample. However, within
%the quasioptical arrangement of the high-field experiments, the
%skin effect  is automatically taken into account by Eq.\
%(\ref{eqtran}) via the complex part of the dielectric
%permittivity, $\varepsilon^*=\varepsilon_1+i\sigma_1/\varepsilon_0
%\omega$. Here $\sigma_1$ is the real part of the complex
%conductivity. Therefore, we can exclude the conductivity (i.e.
%skin-effect) as a reason for the suppression of the line
%intensity. Disorder effects seem also not to play an important
%role, because the line-width remains temperature-independent. The
%physical mechanism of the unexpected temperature behavior of the
%line intensity for $B||b$ remains unclear at present.

The detailed analysis of the AFMR modes in
La$_{0.95}$Sr$_{0.05}$MnO$_3$  and at low temperatures has been
published previously \cite{pimenov00}. In agreement with the
classical predictions for a canted antiferromagnet, two AFMR lines
could be observed, which revealed distinct excitation conditions
and magnetic field-dependence. The full set of the experimental
data including the magnetization and AFMR modes was satisfactorily
explained using a the two-sublattice model of the canted magnetic
structure.

%For completeness, we reproduce in Fig.\ \ref{fig05d} the most
%important result of this work, i.e the field dependence of both
%AFMR modes for different orientation of the magnetic field. It is
%important to note, that the model parameters have been fixed by
%fitting of the anisotropic magnetization data and by the resonance
%position in zero field. Therefore, the model contained no free
%parameter to fit the field and orientation dependencies. Taking
%this in mind, the model curves describe well the observed
%behavior. An interesting point is the indication of the predicted
%spin-flop transition around $B=7.5$\,T (middle panel of Fig.\
%\ref{fig05m}). This transition is smeared probably due to small
%misalignment of the crystal, which is known \cite{hagedorn} to
%strongly smear out the spin-flop transition.

A qualitatively similar behavior has also been observed for
La$_{0.925}$Sr$_{0.075}$MnO$_3$. However, this sample turned out
to be heavily twinned and the separation of the different
orientations was not possible. As will be shown later (Fig.\
\ref{figd}, Sec.\ \ref{secsum}), the field dependence of the
antiferromagnetic resonances is well described by the
two-sublattice model of the canted structure.

\subsection{The Ferromagnetic Insulator ($0.10\le x\le 0.15$)}

The doping dependence of the AFMR-frequencies in the low-$x$
region of the phase diagram, which has been discussed in the
previous sections, reveals a gradual softening of the
quasi-ferromagnetic mode and the weakening of the
quasi-antiferromagnetic mode (Fig. \ref{figsh}). In de Gennes
scenario a ferromagnetic and metallic phase follows the CAF
structure. In the manganites an intermediate ferromagnetic
insulating state is found for Sr and Ca doping. This FM/I most
probably results from a suppression of the static Jahn-Teller
distortions, the increasing importance of the orbital degeneracy
and a subsequent onset of a new orbital order which stabilizes the
ferromagnetic insulator. In the ferromagnetic state
demagnetization effects due to the spontaneous magnetization start
to become important. In addition, the insulating character of the
manganites in this composition range leads to additional
complications of the spectra.

\begin{figure}[]
\centering
\includegraphics[angle=0,width=6cm,clip]{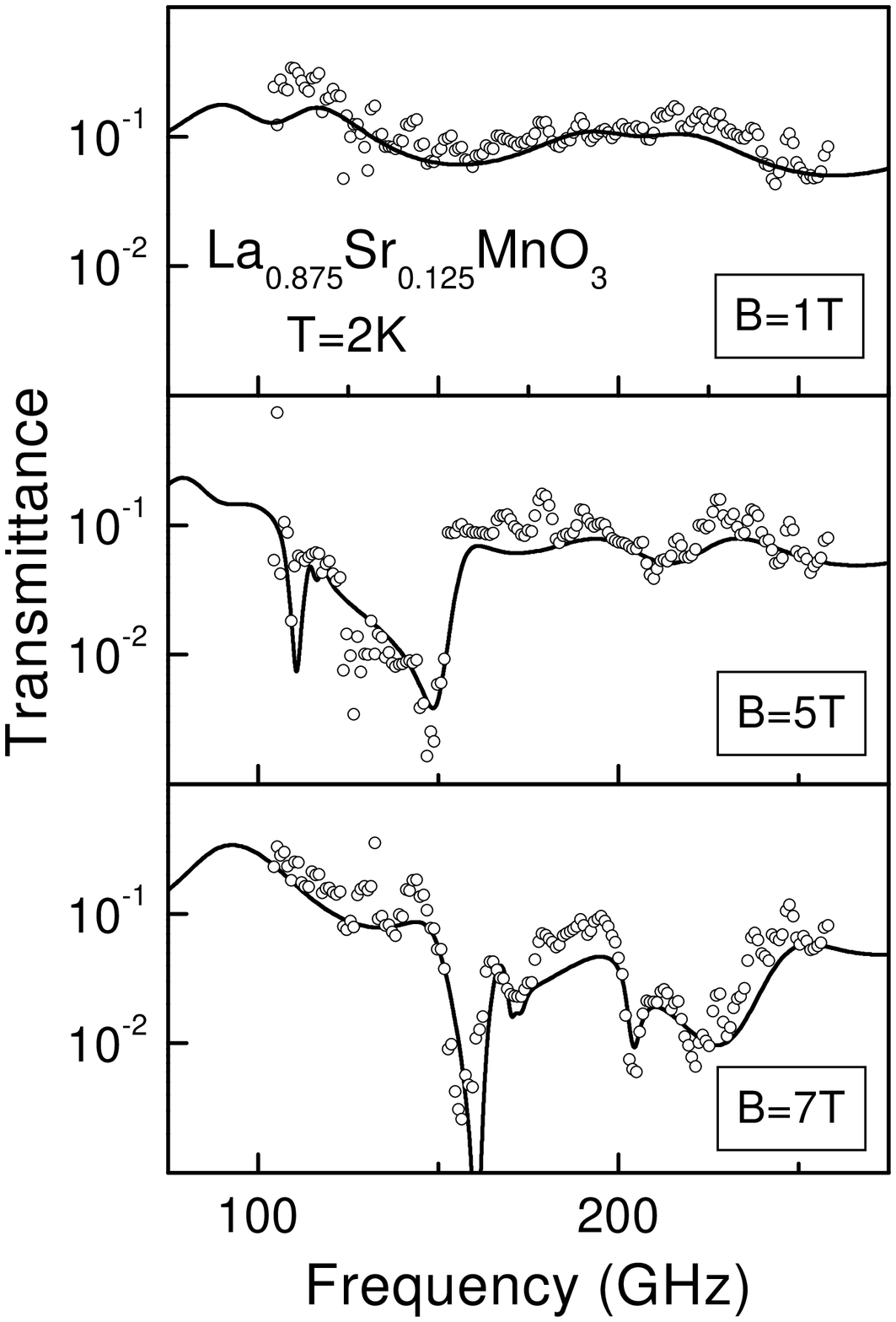}
\vspace{0.2cm} \caption{Frequency-dependent transmittance of a 0.194 mm thick
La$_{0.875}$Sr$_{0.125}$MnO$_{3}$ single crystal in
 Faraday geometry. Upper panel: $B=1\,$T, the FMR mode is below
the frequency range of the experiment. Middle ($B=5\,$T) and lower ($B=7\,$T) panel: the
complicated character of the FMR spectra is due to interference effects and Faraday
rotation. Symbols - experiment. The lines represent the calculated transmittance of the
platelet-shaped sample assuming a parallel orientation of the polarizer and analyzer. The
model parameters are the same for all curves: $g=2,\ M\simeq4\,\mu_{\rm B}$/Mg-atom,
$\varepsilon^*=56+i5.7$, FMR linewidth $\delta=0.1\,$cm$^{-1}$.} \label{fig12f}
\end{figure}

\begin{figure}[]
\centering
\includegraphics[width=6cm,clip]{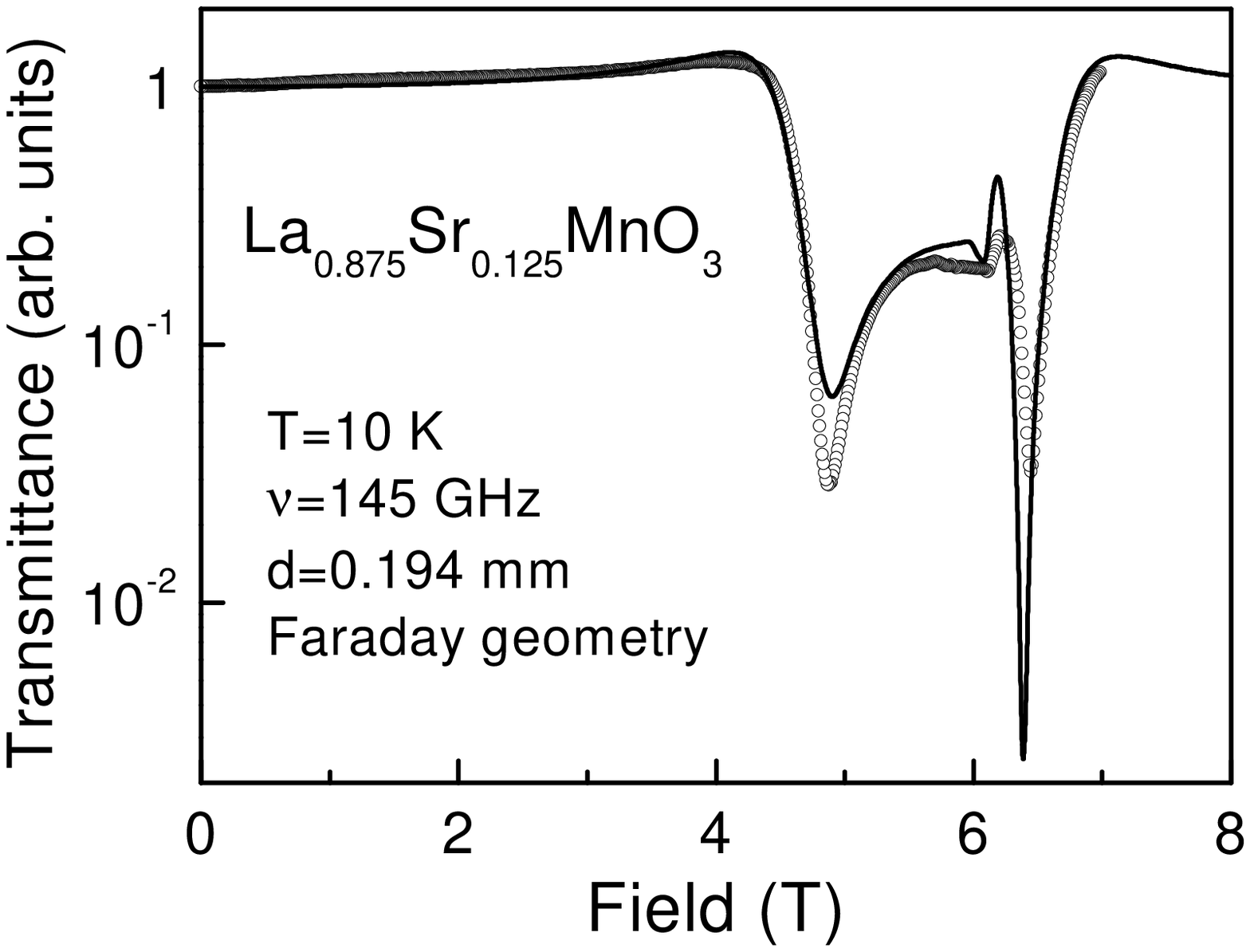}
\vspace{0.2cm} \caption{High-field ESR spectra of La$_{0.875}$Sr$_{0.125}$MnO$_{3}$ in
the Faraday/transmission geometry. Symbols - experiment, line - theory, which takes into
account only a single ferromagnetic resonance mode and electrodynamic effects in the
Faraday geometry (see text). The model parameters are the same as given in Fig.\
\ref{fig12f}.}
 \label{fig12m}
\end{figure}

Figures\ \ref{fig12f} and \ref{fig12m} shows the high-field ESR
spectra of La$_{1-x}$Sr$_x$MnO$_3$ with $x=0.125$ measured in
Faraday geometry. Compared to the spectra of the lower-doped
manganites, these data are more difficult to interpret. We recall
that in this concentration range La$_{1-x}$Sr$_x$MnO$_3$ is a
ferromagnetic insulator and the spectra are expected to consist of
a single FMR mode. The physical origin of the complicated spectra
observed is due to additional electrodynamic effects, which arise
in Faraday geometry and for high intensities of the FMR mode. In
the Faraday geometry left- and right-rotating circular polarized
components of the incident radiation reveal strongly different
propagation constants near resonance. As a result, the linearly
polarized incident wave transforms into an elliptically polarized
wave. Depending on the relative orientation of analyzer and
polarizer qualitatively different transmission (reflection)
spectra are expected. Multiple reflections of the radiation within
the platelet-shaped sample result in additional peculiarities.
Similar effects in the vicinity of a AFMR mode have been observed
previously in YFeO$_3$ orthoferrite \cite{mukhin97}.

Within the transmission geometry of our experiments the analyzer
is positioned parallel to the polarizer. Taking into account the
above-mentioned effects we have simulated these spectra both for
the frequency (Fig. \ref{fig12f}) and field (Fig. \ref{fig12m})
sweeping modes. A single ferromagnetic resonance mode Eqs.
(\ref{eqreson},\,\ref{faraday}) has been taken into account using
$g=2$  and a magnetic susceptibility characteristic for an
ordinary ferromagnet. The FMR linewidth was used as the only
fitting parameter ($\delta=0.1-0.15$\,cm$^{-1}$), since the
intensity and the frequency shift of the FMR mode is directly
determined by the magnetization, which was taken from static
measurements ($M\simeq4\,\mu_{\rm B}$/Mg-atom
\cite{parask00,paraskmmm}).
%(sample density $\rho\simeq 6.5-6.7$\,g/cm$^3$).
The complex dielectric permittivity, which is also important to
explain these phenomena was determined by fits of the
transmittance at $H=0$ as $\varepsilon^*=56+i5.7$. No further
fitting parameters have been utilized. The reasonable agreement
between theory and experiment demonstrates the validity of our
ansatz using a single FMR mode. Similar transmittance spectra were
also observed for the composition $x=0.1$ (not shown), which
revealed even more pronounced peculiarities in a thicker sample.
%Moreover,
%our simulation of the transmission spectra distinctly indicate an
%existence of the single intensive FMR line for the orientation of
%the analyzer for 45 degrees with respect to the polarizer (or -45
%degrees, depending on a magnetic field polarity). In this case the
%line positions both for the frequency and field sweeping modes are
%between the corresponding two interference peaks in Fig. 6a, b.
%
Finally, we note that in some sense the observed peculiarities of
the transmission spectra can be considered as magnetostatic modes
\cite{magneto}.

\begin{figure}[]
\centering
\includegraphics[width=6cm,clip]{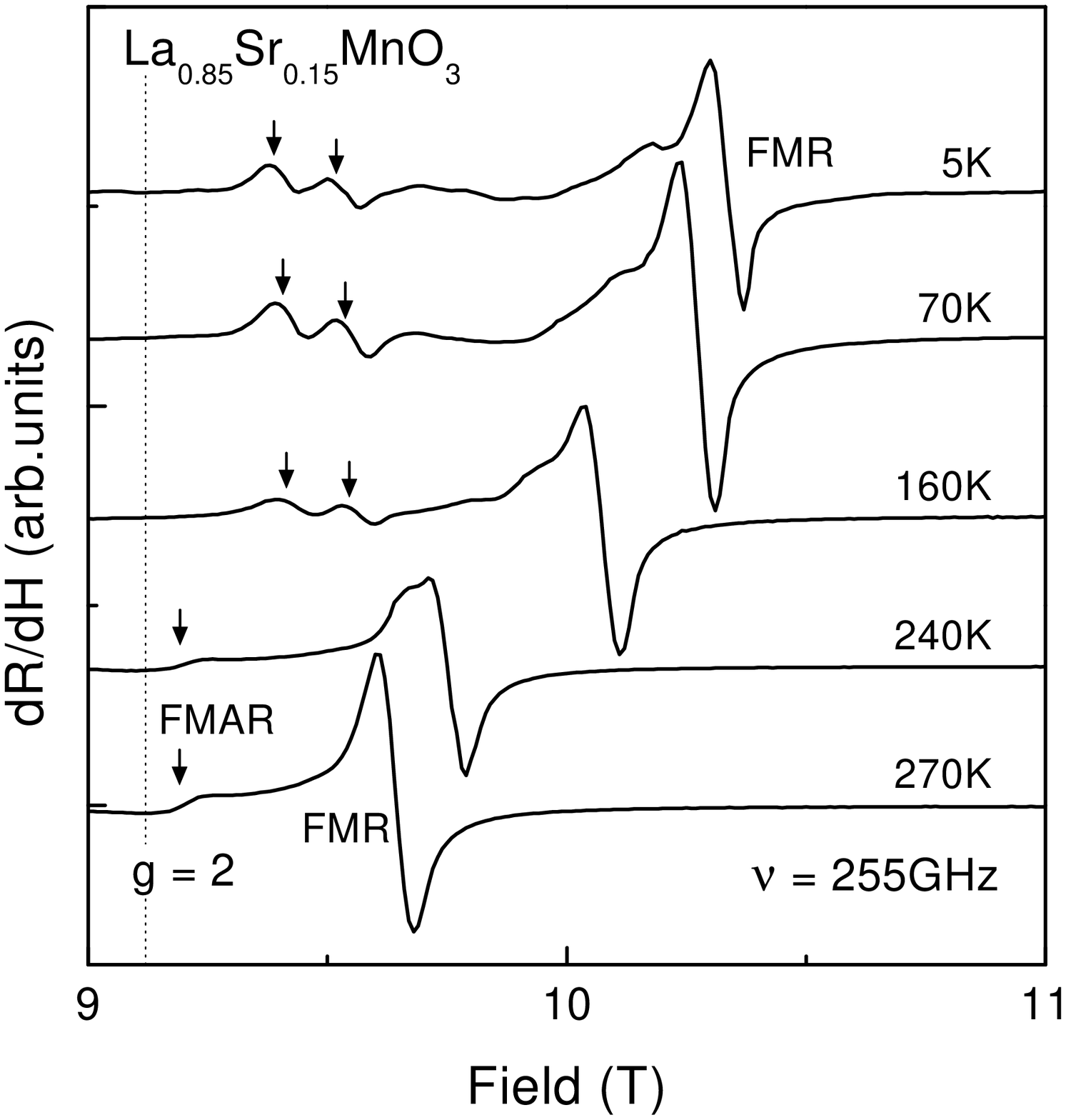}
\vspace{0.2cm} \caption{High-field ESR spectra of La$_{0.85}$Sr$_{0.15}$MnO$_{3}$ in the
Faraday/reflection geometry revealing ferromagnetic resonance and antiresonance modes.
The curves are shifted for clarity. The antiresonance modes are marked by arrows. The
line splitting results from the residual influence of the electrodynamic effects
(magnetostatic modes).} \label{fig15m}
\end{figure}

Closer to the ferromagnetic metallic region of the phase diagram
($x\gtrsim 0.15$) the increase of the conductivity hampers the
transmission experiments. Therefore in this composition range the
reflectance geometry has been employed in our high-field ESR
experiments. A typical example of a field-sweep curve is
represented in Fig.\ \ref{fig15m} showing ESR spectra of
La$_{0.85}$Sr$_{0.15}$MnO$_3$. We recall that the reflection
experiments have been carried out within the Faraday geometry. The
spectra in Fig. \ref{fig15m} are dominated by the single
ferromagnetic resonance (FMR), which is shifted to higher magnetic
fields by the value of the static magnetization (cf. Eq.
\ref{faraday}). Due to only weak interference effects in
reflection measurements of thick samples, the line of the
ferromagnetic \emph{antiresonance} (FMAR) \cite{morrish} becomes
clearly visible and is indicated by arrows.

FMAR line has previously been observed in manganites \cite{wang}
as a minimum in the microwave absorption. The FMAR line
corresponds to the zero crossing of the real part of the magnetic
susceptibility ($\mu^*(\omega,B)=\mu_1+i\mu_2$), which leads to a
minimum in the reflectance. Therefore, in the field derivative
($dR/dH$) this line reveals an opposite sign, compared to the
FMR-mode, which corresponds to a local maximum in reflectance.

\begin{figure}[]
\centering
\includegraphics[width=6cm,clip]{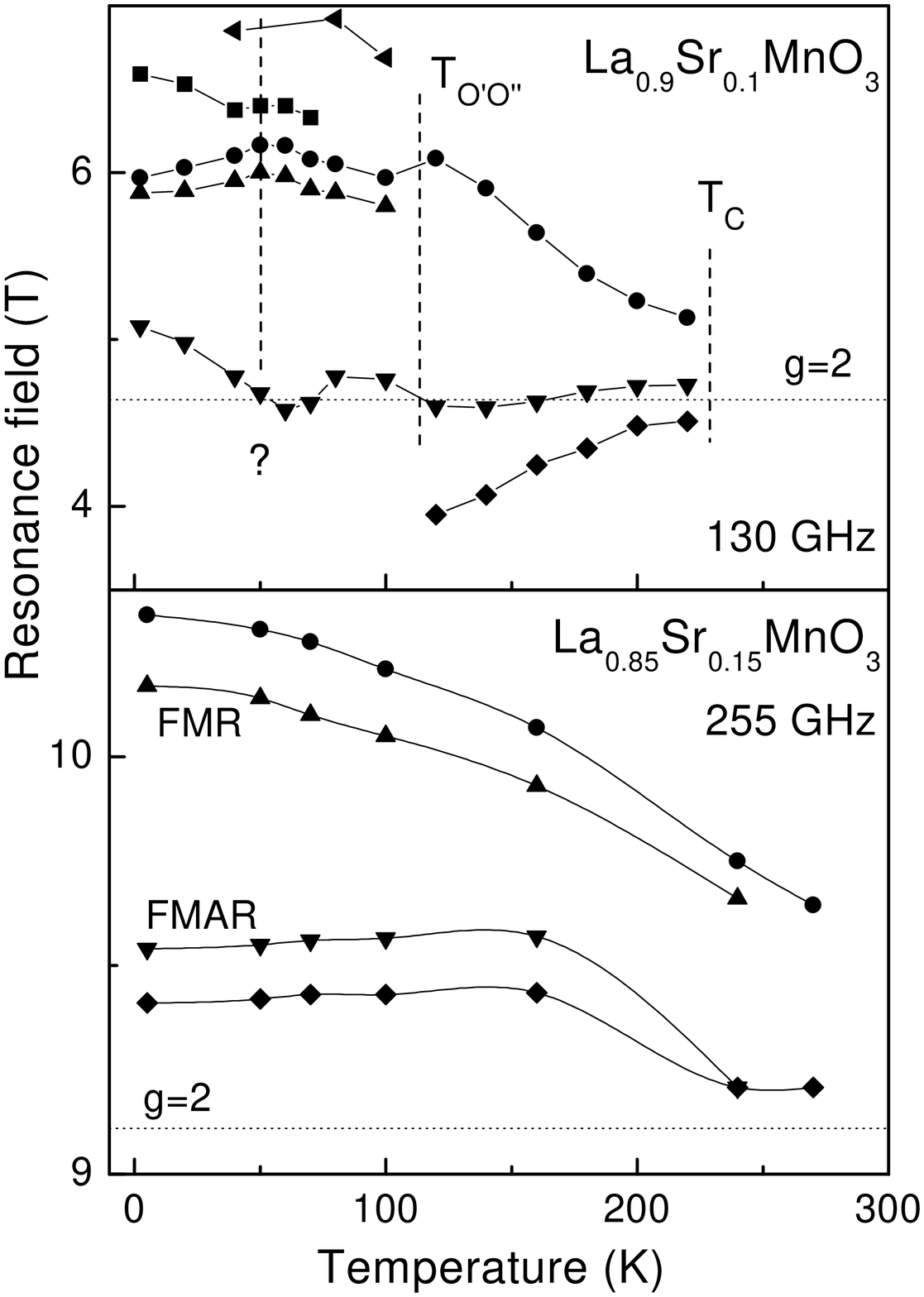}
\vspace{0.2cm} \caption{Temperature dependence of the ESR modes in the Faraday geometry.
Solid lines are drawn to guide the eye.
Symbols represent different observed modes. Dotted lines indicate $g=2$.\\
Upper panel: La$_{0.9}$Sr$_{0.1}$MnO$_{3}$. Dashed lines mark the
phase-transitions temperatures. \\
Lower panel: La$_{0.85}$Sr$_{0.15}$MnO$_{3}$.} \label{fig10d}
\end{figure}

Figure\ \ref{fig10d} shows the temperature dependence of the
high-fiels EPR modes (i.e. positions of the interference minima in
the transmission) for La$_{0.9}$Sr$_{0.1}$MnO$_3$ and
La$_{0.85}$Sr$_{0.15}$MnO$_3$. The complicated pattern observed
for $x=0.1$ and shown in the upper panel is related to the above
discussed interference effects near the ferromagnetic resonance.
The temperature dependence of the line positions reflects the
change of the magnetization and of the dielectric permittivity.
The number of modes becomes reduced for higher temperatures, i.e.
as the sample is getting closer to paramagnetic state where
gyrotropic and interference  effects are much weaker. The data in
the upper panel of Fig.\ \ref{fig10d} correspond to a static
magnetic field of 5-6\,T. This is most probably the reason, why
the magnetic ordering near $210$\,K becomes rather smooth and is
shifted by about 60\,K compared to $T_{\rm C}\simeq 150\,$K
observed in zero magnetic field. In contrary, the transition to
the orbital-ordered state $T_{\rm O'O''}\simeq 120$\,K is only
weakly affected by the magnetic field. In addition to these well
known transitions, the temperature dependence of the ESR lines
reveal an anomaly around 60\,K. The nature of this transition is
not clear at present. In this context we would like to mention the
recent observation \cite{zhou01} of anomalies in the thermal
conductivity of La$_{0.9}$Sr$_{0.1}$MnO$_3$ around $T_{\rm
LO}\simeq 75$\,K, which may correspond to the anomalies in the
high-field ESR spectra, observed around $T=60$\,K.

The lower panel of Fig.\ \ref{fig10d} shows the temperature
dependence of the FMR and FMAR modes of
La$_{0.85}$Sr$_{0.15}$MnO$_3$. The observed line splitting is
probably due to interference effects and to different g-factors
for non-equivalent Mn sites. The frequency of the ESR experiment
approximately corresponds   to a static magnetic field of 10\,T.
Such a high field field smears the magnetic-ordering transition
and strongly increases the transition temperature for
La$_{0.85}$Sr$_{0.15}$MnO$_3$, which can be estimated as $T_{\rm
C}\simeq 350$\,K. This corresponds to a shift of 130\,K compared
to the zero-field value of $T_{\rm C}\simeq 220$\,K
\cite{parask00}.

\subsection{The Ferromagnetic Metal, $x=0.175$}

\begin{figure}[]
\centering
\includegraphics[width=6cm,clip]{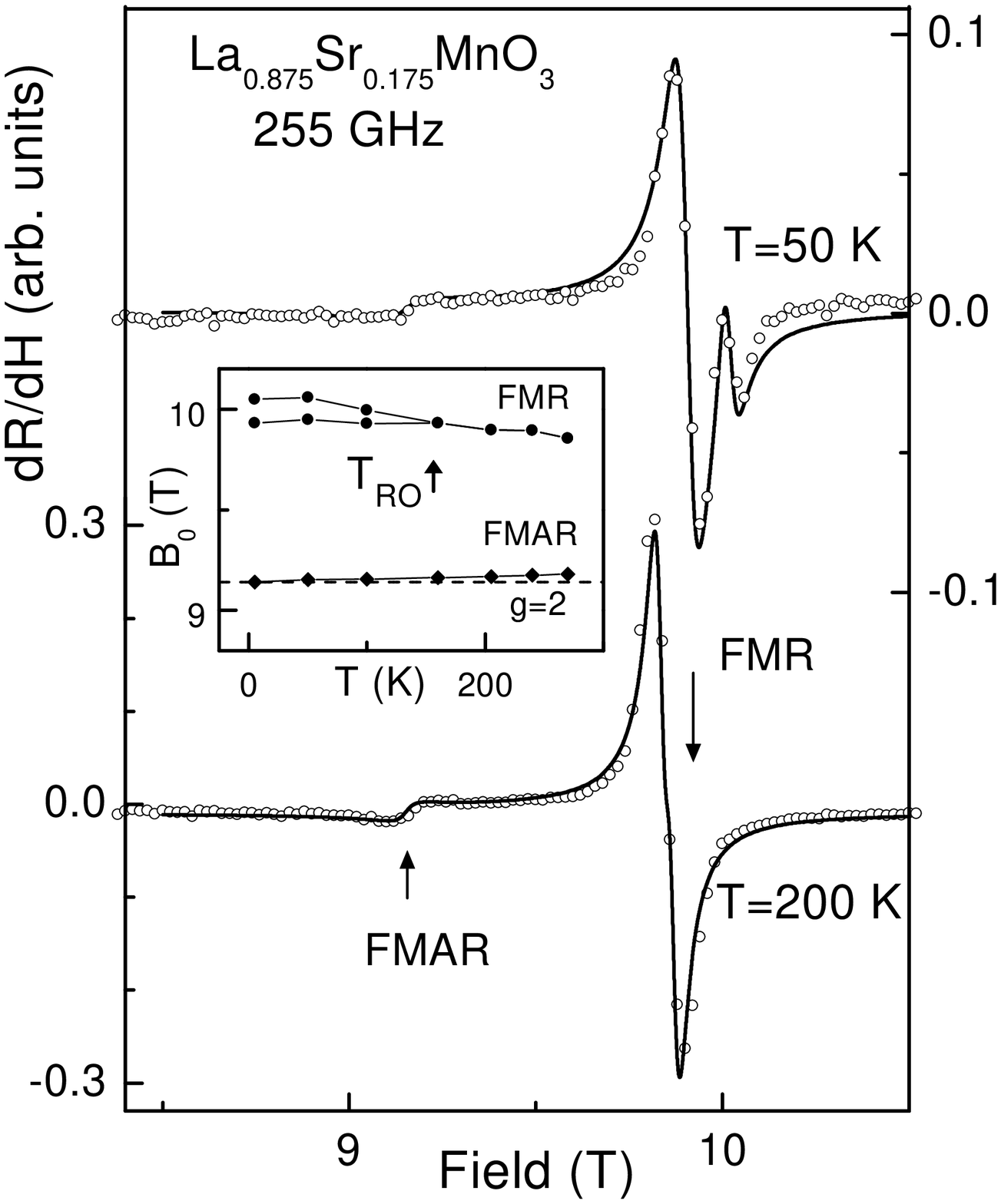}
\vspace{0.2cm} \caption{High-field ESR spectra of La$_{0.825}$Sr$_{0.175}$MnO$_{3}$ in
the Faraday/reflection geometry. The ferromagnetic resonance and antiresonance modes are
marked by arrows. Symbols - experiment, lines - fit. The line splitting of the FMR mode
is due to the anisotropy of the crystal field below the rhombohedral-to-orthorhombic
phase transition at $T_{\rm RO}$. The inset shows the temperature dependence of the
resonance-line positions.} \label{fig17m}
\end{figure}

For $x\geq 0.15$ the conductivity of La$_{1-x}$Sr$_x$MnO$_3$
rapidly increases  resulting in a significant increase of the
absolute reflection. Typical high-field ESR spectra in this
composition range are represented in Fig.\ \ref{fig17m}. As
observed in La$_{0.85}$Sr$_{0.15}$MnO$_3$, the spectra obtained in
La$_{0.825}$Sr$_{0.175}$MnO$_3$ show two lines of opposite sign,
which correspond to FMR and FMAR modes. Because
La$_{0.825}$Sr$_{0.175}$MnO$_3$ is already in the ferromagnetic
state at all temperatures investigated, both lines reveal no
substantial temperature dependence. The intensity of these lines
decreases for decreasing temperature. This effect is easily
explained by the increase of the conductivity towards low
temperatures, and, consequently, by the increase of the absolute
reflectance. In this case the relative effect of the magnetic
resonance line decreases. Assuming a temperature independent
contribution of the FMR mode the simulation of the line intensity
yields a value of the dc-resistivity, $\rho_{dc}=1/\sigma_1\sim
10^{-4}\,\Omega\cdot$cm at low temperature, which is in agreement
with the measured value \cite{mukhin98}.

At low temperatures the line of the ferromagnetic resonance
becomes splitted. The analysis of the temperature dependence of
the splitting (inset of Fig.\ \ref{fig17m}) shows that the
splitting starts at the rhombohedral-to-orthorhombic (R/O)
transition \cite{parask00,mukhin98,liu}. Using the
temperature-dependent powder X-ray spectroscopy, the R/O
transition temperature has been determined as $T_{\rm RO}\simeq
180$\,K for this sample. Therefore, we attribute the splitting of
the FMR line to the increase of the crystal anisotropy at the R/O
transition. (The simultaneous observation of two lines arises from
the twinning of the sample.) This effect is directly connected to
lowering of the crystal symmetry below $T_{\rm RO}$. In principle,
one could expect a similar splitting for the FMR lines for $0.1
\le x\le 0.15$, which are all in the orthorhombically distorted
phase (O$''$). But this effect is probably masked by the larger
linewidth for these concentrations. We note that the interference
effects are not excluded completely even for a highly conducting
La$_{0.825}$Sr$_{0.175}$MnO$_3$.
%Comparing the
%ESR-line positions for x=0.15 and 0.175 compositions a significant
%($\sim 3-4\%$) smaller g-factor value for x=0.15 composition as
%compare to x=0.175 one. (It could be related to the differen
%character (metallic and insulating) of the ground electronic
%states in these compounds.

\section{Summary \label{secsum}}

\begin{widetext}
\begin{figure}[]
%\centering
\includegraphics[angle=270,width=16cm,clip]{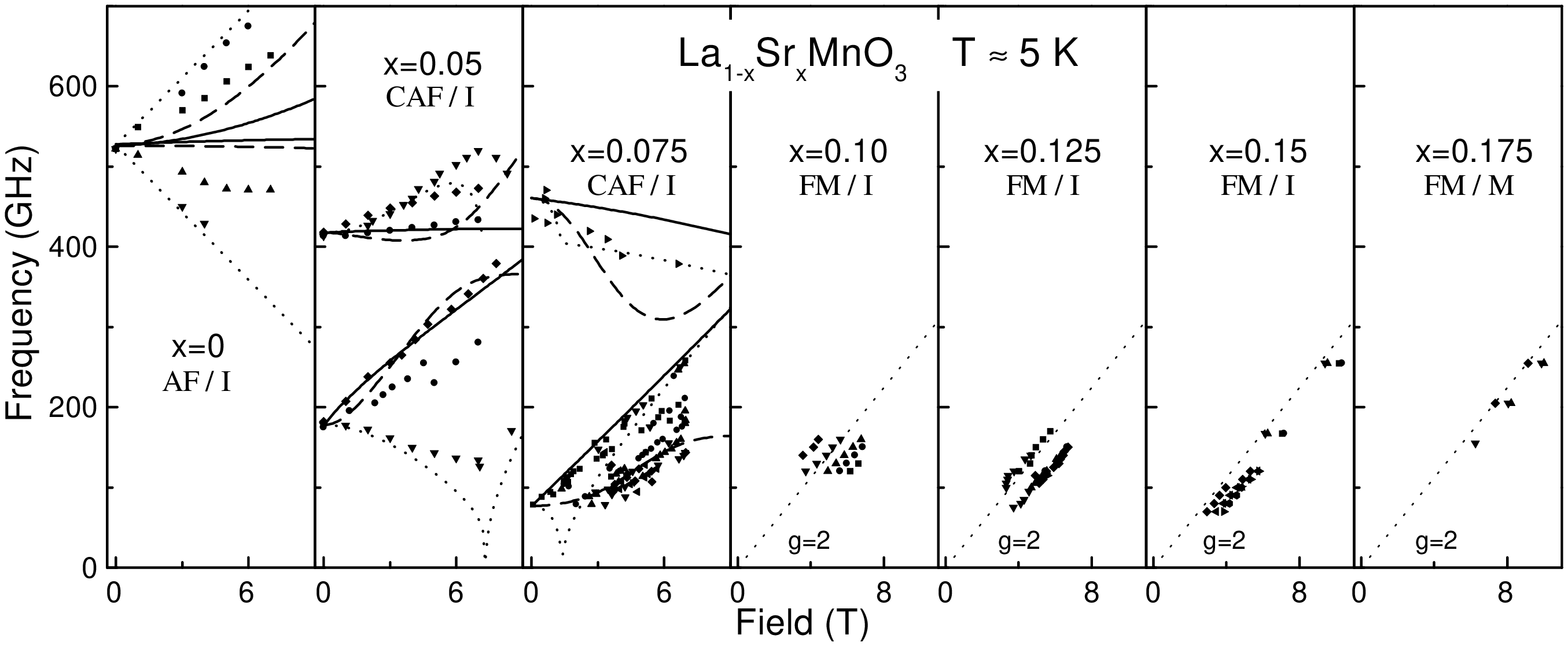}
\caption{ Summary of the high-field ESR spectra of La$_{1-x}$Sr$_{x}$MnO$_{3}$ at low
temperatures. AF - antiferromagnet, CAF - canted AF, FM - ferromagnet, I - insulator, M -
metal. Symbols - experiment. Lines for $x \le 0.075$ represent the model for the canted
magnetic structure and the $g=2$-position for $x \geq 0.1$. The sample orientation and
excitation conditions are determined unambiguously for the $x=0.05$ concentration only
(see Ref. \cite{pimenov00} for details). The data for $x \geq 0.1$ have been obtained in
the Faraday geometry.} \label{figd}
\end{figure}
\end{widetext}

 Figure\ \ref{figd} summarizes the
high-field ESR results in La$_{1-x}$Sr$_x$MnO$_3$ for all
 compositions investigated:

The parent compound LaMnO$_3$ is purely antiferromagnetic and
reveals two nearly degenerate AFMR lines ($x=0$). These lines
become splitted in an external magnetic field, as it is expected
for a conventional antiferromagnet.

In the lowest-doping regime  $0\le x\le 0.075$ the compounds reveal close similarities
concerning the appearance and the splitting of the AFMR modes. Already without magnetic
field for increasing Sr-concentration the initially degenerate AFMR modes split into
quasi-ferromagnetic (F) and quasi-antiferromagnetic (AF) modes. The F-mode rapidly
softens and finally transforms into the line of ferromagnetic resonance for $x\gtrsim
0.09$. The AF-mode reveals a much weaker composition dependence and disappears in the
ferromagnetic state. The lines in Fig.\ \ref{figd} for $0\le x\le 0.075$ represent the
calculations using the two-sublattice model of the canted magnetic structure (Section
\ref{seccan}). Fig.\ \ref{figd} shows the results for all orientations of the magnetic
field simultaneously. For $x=0$ and $x=0.075$ this cannot be avoided

\noindent because of the twinning of the samples. For $x=0.05$ the comparison between
theory and experiment can be carried out for all orientations separately (see Ref.\
\cite{pimenov00} for a detailed analysis). Taking into account the twinning of the
crystals, the two-sublattice model of the canted magnetic structure can well reproduce
the high-field EPR spectra of the low-doped La$_{1-x}$Sr$_x$MnO$_3$ compounds.

The composition range $0.1\le x\le 0.15$ corresponds to the
ferromagnetic insulator at low temperatures. The ESR spectra in
this concentration range are rather complicated, which can be
attributed to the interference of two normal modes (right- and
left-circular polarizations) near the ferromagnetic resonance,
resulting in a significant change of the polarization of the
incident radiation.  The spectra in Fig.\ \ref{figd} are shown for
the Faraday geometry and are therefore shifted to higher fields
compared to $g=2$ (dashed lines for $0.1\le x\le 0.175$). All data
can be satisfactorily accounted for using a \emph{single} FMR
mode.

The high-field ESR spectra for higher doping levels show
ferromagnetic resonance (FMR) and ferromagnetic antiresonance
(FMAR) modes. In addition, a splitting of the FMR mode is observed
in La$_{0.825}$Sr$_{0.175}$MnO$_3$ below the structural
orthorhombic/rhombohedral transition which is attributed to the
lowering of the crystal symmetry.

\begin{figure}[b]
%\centering
\includegraphics[width=6cm,clip]{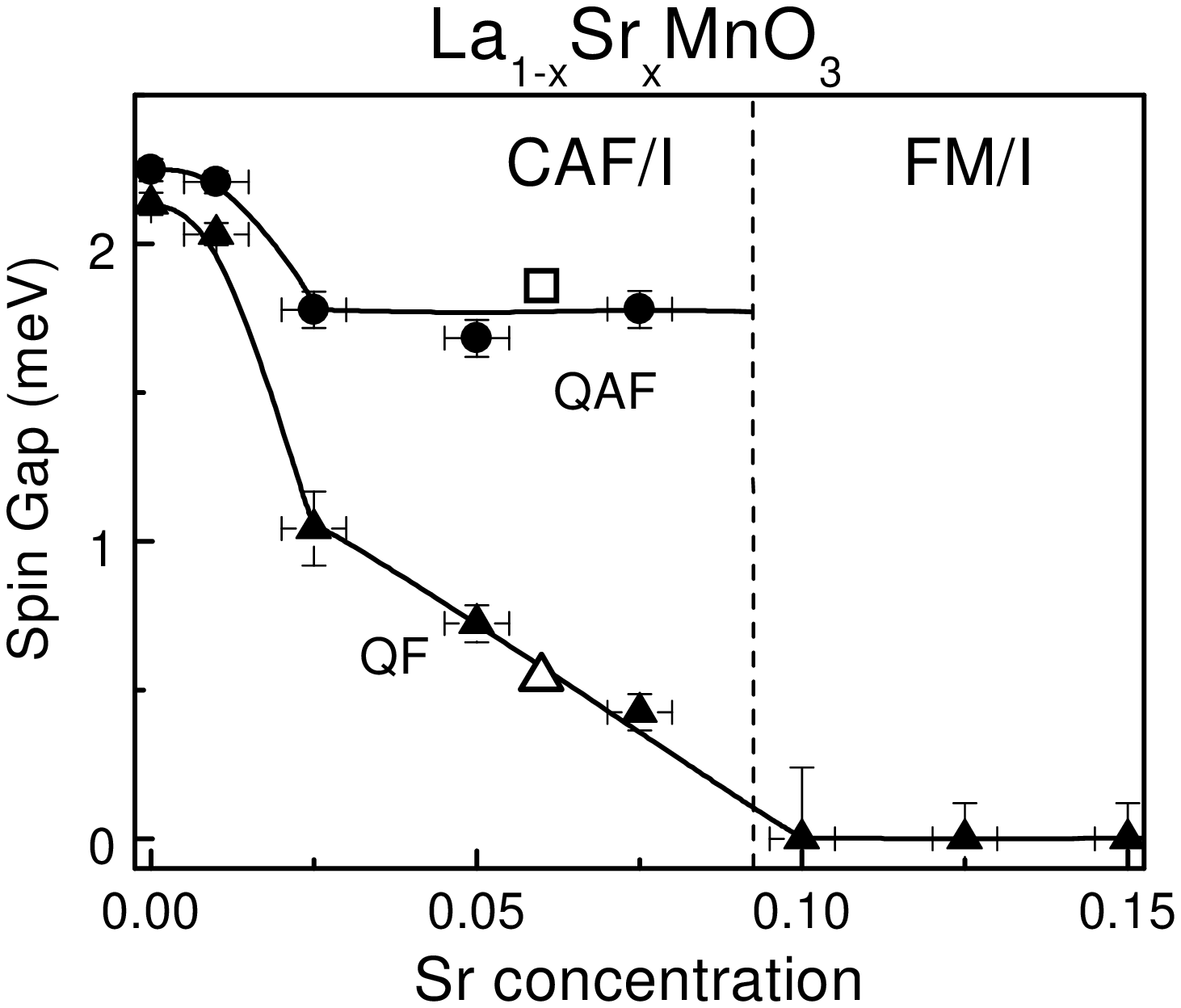}
\vspace{0.2cm} \caption{Doping dependence of the spin gap in La$_{1-x}$Sr$_x$MnO$_3$ from
AFMR \cite{mukhin00} and high-field ESR experiments. Closed circles -
quasi-antiferromagnetic (QAF) mode, closed triangles - quasi-ferromagnetic (QF) mode (see
Sec. \ref{seccan} for details). Open symbols represent the neutron-scattering results
\cite{hennion00}. Note the disappearance of the QAF branch at $x\sim 0.1$. The unusual
doping dependence around $x\sim 0.025$ reflects a possible threshold concentration
between the antiferromagnetic and canted structures.} \label{figfr}
\end{figure}

Finally, we would like to compare the gap value of the spin-wave
branches as determined in the present work with those as derived
from neutron-scattering experiments in Sr-doped manganites. The
wavelength of the submillimeter radiation is much larger than the
interatomic distance. Therefore the high-field ESR determines the
gap values at the center of the Brillouin zone. The concentration
dependence of the high-frequency (QAF) and low-frequency (QF) mode
are shown in Fig.\ \ref{figfr} as determined at $T\simeq5$\,K.

The high-field ESR results agree well with the neutron-scattering
data for La$_{0.94}$Sr$_{0.06}$MnO$_3$ \cite{hennion00}, which are
included in Fig.\ \ref{figfr} for comparison. For $0.025 \le x\le
0.1$ the concentration behavior of the quasi-ferromagnetic branch
is approximately linear. On the contrary, the
quasi-antiferromagnetic mode is roughly constant in this
concentration range. At $x\simeq 0.1$ the low-frequency branch
becomes zero and the high-frequency mode disappears. This behavior
corresponds well to the transition from the CAF to the
ferromagnetic state with a gapless spin-wave branch. A very small
nonzero gap ($\nu_0\sim 0.019\,$meV) has been observed in
La$_{0.85}$Sr$_{0.15}$MnO$_3$ in inelastic neutron-scattering
experiments \cite{doloc}. However, this value cannot be
distinguished from zero within the accuracy of our experiment.
Nevertheless, the splitting of the FMR line for $x=0.175$ is of
comparable amplitude (0.12\,T$\hat{=}0.014$\,meV). Finally, we
note that for $x\geq 0.1$ a FM ground state with a different
orbital order is established.

An unusual behavior of the AFMR frequencies is observed for low
concentrations (Fig.\ \ref{figfr}). For $x \le 0.025$ both modes
are close to each other and approximately independent on doping.
We suggest that this may reflect the stability of the pure
antiferromagnetic state against canting for low doping
concentration. Above a threshold value of $x\simeq 0.025$ the
doping by Sr$^{2+}$ leads to the abrupt increase of the canting
angle and to the corresponding drop of the AFMR frequencies. We
note that such a transition above some critical concentration has
been predicted theoretically by several authors \cite{nagaev79}.

In conclusion, using the high-field ESR technique we have
investigated the magnetic properties of La$_{1-x}$Sr$_x$MnO$_3$
for the composition range from antiferromagnetic insulator up to
ferromagnetic metal. In undoped LaMnO$_3$ a splitting of the
antiferromagnetic resonance mode is observed only in the external
magnetic field in agreement with the antiferromagnetic structure
of this material. For increasing Sr-doping the AFMR modes are
splitted even in zero field, which can be naturally explained
assuming a canted magnetic structure for $x<0.1$. A possible
threshold concentration from the antiferromagnetic to the canted
state is observed around $x\simeq 0.025$. In the ferromagnetic
insulating state ($0.1\le x \le 0.15$) a complicated set of
spectra has been observed. However, these spectra can be well
explained by a single ferromagnetic resonance mode and taking into
account electrodynamic effects. In La$_{0.825}$Sr$_{0.175}$MnO$_3$
the ESR spectra are consistent with the ferromagnetic and metallic
state. The lines of ferromagnetic resonance and  ferromagnetic
antiresonance can be clearly observed in reflectance geometry.
Therefore, the ground state of La$_{1-x}$Sr$_x$MnO$_3$ for $x\le
0.175$ can be well characterized by the high-field ESR technique.

\section{Acknowledgements}

We acknowledge stimulating discussion with M. Hennion, D. Khomski,
J. Deisenhofer and M. Paraskevopoulos. We are indebted to M.
M\"{u}ller, A. Pimenova and F. Mayr for carrying out SQUID and
X-ray experiments. This work was supported in part by BMBF
(13N6917/0 - EKM), by DFG (SFB 484), by INTAS (97-30850) and RFBR
(99-02-16848).


\begin{thebibliography}{99}
\bibitem{jonker}  G. H. Jonker and J. H. van Santen, Physica {\bf 16}, 337
(1950); {\bf 19}, 120 (1953)

\bibitem{wollan}  E. O. Wollan and W. C. Koehler, Phys. Rev. {\bf 100}, 545
(1955).

\bibitem{degennes}  P.-G. de Gennes, Phys. Rev. {\bf 118}, 141 (1960).

\bibitem{zener}  C. Zener, Phys. Rev. {\bf 82}, 403 (1951).

\bibitem{gmr}  K. Chahara, T. Ohno, M. Kasai and Y. Kozono, Appl. Phys.
Lett. {\bf 63}, 1990 (1993); R. von Helmolt, J. Wecker, B.
Holzapfel, L. Schulz and K. Samwer, Phys. Rev. Lett. {\bf 71},
2331 (1993); S. Jin, T. H. Tiefel, M. McCormack, R. A. Fastnacht,
R. Ramesh and L. H. Chen, Science {\bf 264}, 413 (1994).

\bibitem{furukawa}  N. Furukawa, J. Phys. Soc. Jpn. {\bf 63}, 3214 (1994).

\bibitem{millis95}  A. J. Millis, P. B. Littlewood and B. I. Shraiman,
Phys. Rev. Lett. {\bf 74}, 5144 (1995).

\bibitem{roder} H. R\"{o}der, Jun Zang, and A. R. Bishop, Phys. Rev. Lett. \textbf{76}, 1356
(1996).

\bibitem{moreo99} A. Moreo, S. Yunoki, and E.
Dagotto, Science \textbf{283}, 2034 (1999); A. Moreo, M. Mayr, A.
Feiguin, S. Yunoki, and E. Dagotto, Phys. Rev. Lett. \textbf{84},
5568 (2000).

\bibitem{maezono} R. Maezono, S. Ishihara, and N. Nagaosa, Phys.
Rev. B \textbf{58}, 11583 (1998).

\bibitem{kilian} R. Kilian and G. Khaliullin, Phys. Rev. B \textbf{58},
11841 (1998).

\bibitem{horsch} P. Horsch, J. Jakli\v{c}, and F. Mack, Phys. Rev. B
\textbf{59}, 6217 (1999).

\bibitem{brink} J. van den Brink, P. Horsch, F. Mack, and A. M.
Ole\'{s}, Phys. Rev. B 59, 6795 (1999).

\bibitem{jirak} Z. Jir\'{a}k, S. Krupi\v{c}ka, Z. \v{S}im\v{s}a,
M. Dlouh\'{a}, and S. Vratislav, J. Magn. Magn. Mater.
\textbf{53}, 153 (1985).

\bibitem{mukhin00}  A. A. Mukhin, V. Yu. Ivanov, V. D. Travkin, A. Pimenov,
A. Loidl, and A. M. Balbashov, Europhys. Lett., {\bf 49}, 514
(2000).

\bibitem{pimenov00} A. Pimenov, M. Biberacher, D. Ivannikov, A. Loidl, V. Yu. Ivanov,
 A. A. Mukhin, and A. M. Balbashov, Phys. Rev. B \textbf{62}, 5685 (2000).

\bibitem{hennion97} M. Hennion, F. Moussa, J. Rodr\'{\i}guez-Carvajal, L. Pinsard,
and A. Revcolevschi, Phys. Rev. B \textbf{56}, R497 (1997).

\bibitem{kawano}  H. Kawano, R. Kajimoto, M. Kubota and H. Yoshizawa, Phys.
Rev. B {\bf 53,} 2202 (1996); H. Kawano, R. Kajimoto, M. Kubota
and H. Yoshizawa, Phys. Rev. B {\bf 53}, R14709 (1996); H.
Yoshizawa, H. Kawano, Y. Tomioka, and Y. Tokura, Phys. Rev. B
\textbf{52}, R13145 (1995).

\bibitem{maiti} A. Chakraborty, D. Bhattacharya, and H. S. Maiti, Phys. Rev. B \textbf{56}, 8828
(1997).

\bibitem{okimoto} Y. Okimoto, T. Katsufuji, T. Ishikawa, T. Arima, and Y.
Tokura, Phys. Rev. B \textbf{55}, 4206 (1997).

\bibitem{brion} S. de Brion, F. Ciorcas, G. Chouteau, P. Lejay, P. Radaelli, and C.
Chaillout, Phys. Rev. B \textbf{59}, 1304 (1999).

\bibitem{skumryev} V. Skumryev, F. Ott, J.M.D. Coey, A. Anane, J.-P. Renard,
L. Pinsard-Gaudart, and A. Revcolevschi, Eur. Phys. J. B
\textbf{11}, 401 (1999).

\bibitem{nagaev}  E. L. Nagaev, Sov. Phys.- Uspekhi {\bf 39}, 781
(1996).

\bibitem{nagaev98}  E. L. Nagaev, Phys. Rev. B {\bf 58}, 2415
(1998).

\bibitem{yunoki}  S. Yunoki, J. Hu, A. L. Malvezzi, A. Moreo, N. Furukawa,
and E. Dagotto, Phys. Rev. Lett. {\bf 80}, 845 (1998); S. Yunoki,
A. Moreo, and E. Dagotto, Phys. Rev. Lett. {\bf 81}, 5612 (1998).

\bibitem{dagotto01} E. Dagotto, T. Hotta, and A. Moreo, Phys. Rep.
\textbf{344}, 1 (2001).

\bibitem{kagan01} M. Yu. Kagan and K. I. Kugel', Physics-Uspekhi
\textbf{44}, 553 (2001).

\bibitem{solovyev96} I. Solovyev, N. Hamada, and K. Terakura, Phys. Rev. Lett. \textbf{76}, 4825
(1996).

\bibitem{arovas} D. P. Arovas and F. Guinea, Phys. Rev. B \textbf{58}, 9150
(1998).

\bibitem{solovyev01} I. V. Solovyev and K. Terakura, Phys. Rev. B
\textbf{63}, 174425 (2001)

\bibitem{zou} L.-J. Zou, H. Q. Lin, and D. K. Campbell, Phys. Rev. B \textbf{63}, 214402
(2001).

\bibitem{mori} S. Mori, C. H. Chen, and S.-W. Cheong, Nature \textbf{392},
473 (1998).

\bibitem{radaelli} P. G. Radaelli, R. M. Ibberson, D. N. Argyriou, H. Casalta,
K. H. Andersen, S.-W. Cheong, and J. F. Mitchell, Phys. Rev. B
\textbf{63}, 172419 (2001).

\bibitem{uehara} M. Uehara, S. Mori, C. H. Chen, and S.-W. Cheong, Nature \textbf{399},
560 (1999).

\bibitem{fath} M. F\"{a}th, S. Freisem, A. A. Menovsky, Y. Tomioka, J. Aarts, and J. A.
Mydosh, Science \textbf{285}, 1540 (1999).

\bibitem{parask00} M. Paraskevopoulos, F. Mayr, J. Hemberger, A.
Loidl, R. Heichele, D. Maurer, V. M\"{u}ller, A. A. Mukhin, and A.
M. Balbashov, J. Phys.: Condens. Matt. \textbf{12}, 3993 (2000).

\bibitem{liu} G.-L. Liu, J.-S. Zhou, and J. B. Goodenough, Phys. Rev. B \textbf{64}, 144414
(2001).

\bibitem{hennion00} M. Hennion, F. Moussa, G. Biotteau, J. Rodr\'{\i}guez-Carvajal,
 L. Pinsard, and A. Revcolevschi, Phys. Rev. B
{\bf 61}, 9513 (2000).

\bibitem{hennion01} G. Biotteau, M. Hennion, F. Moussa, J. Rodr\'{\i}guez-Carvajal, L. Pinsard,
A. Revcolevschi, Y. M. Mukovskii, and D. Shulyatev, Phys. Rev. B
\textbf{64}, 104421 (2001).

\bibitem{hennion98}  M. Hennion, F. Moussa, G. Biotteau, J. Rodr\'{\i}guez-Carvajal, L. Pinsard,
and A. Revcolevschi, Phys. Rev. Lett. {\bf 81}, 1957 (1998).

\bibitem{allodi97}  G. Allodi, R. De Renzi, G. Guidi, F. Licci, and M. W.
Pieper, Phys. Rev. B {\bf 56}, 6036 (1997).

\bibitem{allodi98}  G. Allodi, R. De
Renzi, and G. Guidi, Phys. Rev. B {\bf 57}, 1024 (1998).

\bibitem{kumagai} K. Kumagai, A. Iwai, Y. Tomioka, H. Kuwahara, Y. Tokura, and A.
Yakubovskii, Phys. Rev. B \textbf{59}, 97 (1999).

\bibitem{savosta} M. M. Savosta, P. Nov\'{a}k, M. Mary\v{s}ko, Z. Jir\'{a}k, J. Hejtm\'{a}nek,
J. Englich, J. Kohout, C. Martin, and B. Raveau, Phys. Rev. B
\textbf{62}, 9532 (2000).

\bibitem{roman} J. M. Rom\'{a}n and J. Soto, Phys. Rev. B \textbf{62}, 3300
(2000).

\bibitem{geck} J. Geck, B. B\"{u}chner, M. H\"{u}cker, R. Klingeler, R. Gross,
L. Pinsard-Gaudart, and A. Revcolevschi, Phys. Rev. B \textbf{64},
144430 (2001).

\bibitem{preparation}  A. M. Balbashov, S. G. Karabashev, Ya. M. Mukovskii,
and S. A. Zverkov, J. of Cryst. Growth, {\bf 167}, 365 (1996).

\bibitem{mukhin98}  A. A. Mukhin, V. Yu. Ivanov, V. D. Travkin, S. P. Lebedev,
A. Pimenov, A. Loidl and A. M. Balbashov, JETP Lett., {\bf 68},
356 (1998).

\bibitem{urushibara}  A. Urushibara, Y. Moritomo, T. Arima, A. Asamitsu, G.
Kido, and Y. Tokura, Phys. Rev. B {\bf 51}, 14103 (1995).

\bibitem{moritomo} Y. Moritomo, A. Asamitsu, and Y. Tokura, Phys. Rev. B \textbf{56}, 12190
(1997).

\bibitem{pimenov99} A. Pimenov, Ch. Hartinger, A. Loidl, A. A. Mukhin, V. Yu. Ivanov, A. M.
Balbashov, Phys. Rev. B \textbf{59}, 12419 (1999).

\bibitem{paraskmmm} M. Paraskevopoulos, F. Mayr, C. Hartinger, A. Pimenov,
J. Hemberger, P. Lunkenheimer, A. Loidl, A. A. Mukhin, V. Yu.
Ivanov, and A. M. Balbashov, J. Magn. Magn. Mater. \textbf{211},
118 (2000).

\bibitem{volkov}  G. V. Kozlov and A. A. Volkov in {\it Millimeter and
Submillimeter Wave Spectroscopy of Solids}, Ed. by G. Gr\"{u}ner
(Springer, Berlin, 1998), p. 51.; A. A. Volkov, Yu. G. Goncharov,
G. V. Kozlov, S. P. Lebedev, and A. M. Prochorov, Infrared Phys.
{\bf 25}, 369 (1985).

\bibitem{dupont}  F. Dupont, F. Millange, S. de Brion, A. J\'{a}nossy, and G.
Chouteau, cond-mat/0110139.

\bibitem{zvezdin} A. K. Zvezdin and V. A. Kotov, {\it Modern magnetooptics and magnetooptical
materials} (Inst. of Physics Publ., Bristol, 1997).

\bibitem{born}  M. Born and E. Wolf, {\it Principles of optics} (Pergamon,
Oxford, 1986).

\bibitem{bogush}  A. K. Bogush, V. I. Pavlov and L. V. Balyko, Cryst. Res.
Technol. {\bf 18}, 589 (1983).

\bibitem{yamada}  Y. Yamada, O. Hino, S. Nohdo and R. Kanao, Phys. Rev. Lett.
{\bf 77}, 904 (1996).

\bibitem{zhou}  J.-S. Zhou, J. B. Goodenough, A. Asamitsu and Y. Tokura,
Phys. Rev. Lett. {\bf 79}, 3234 (1997).

\bibitem{matsumoto} G. Matsumoto, J. Phys. Soc. Japan \textbf{29}, 606
(1970).

\bibitem{topfer}  J. T\"{o}pfer and J. B. Goodenough, J. Solid State Chem.
{\bf 130}, 117 (1997).

\bibitem{foner} S. Foner, \textit{Antiferromagnetic and Ferromagnetic
Resonance}, in \textit{Magnetism}, vol. I, edited by G. T. Rado
and H. Suhl (Acad. Press., New York, 1963) p. 383.

\bibitem{turov} E. A. Turov, \textit{Physical Properties of Magnetically
Ordered Crystals}, (Acad. Press., New York, 1965).

\bibitem{mitsido}  S. Mitsudo, K. Hirano, H. Nojiri, M. Motokawa, K. Hirota,
A. Nishizawa, N. Kaneko, and Y. Endoh, J. Magn. Magn. Mater. {\bf
177-181}, 877 (1998).

\bibitem{moria}  T. Morya, in {\it Magnetism}, edited by G. T. Rado and H.
Suhl, Vol. I (Academic Press, NY, 1984) p. 85.

\bibitem{herrmann}  G. F. Herrmann, J. Phys. Chem. Sol. {\bf 24}, 597
(1963); Phys. Rev. {\bf 133}, A1334 (1964).

\bibitem{cinader} G. Cinader, Phys. Rev. \textbf{155}, 453 (1967).

\bibitem{mukhin}  A. A. Mukhin \textit{et al.}, to be published.

\bibitem{ivanshin} V. A. Ivanshin, J. Deisenhofer, H.-A. Krug von Nidda, A. Loidl,
A. A. Mukhin, A. M. Balbashov, and  M. V. Eremin, Phys. Rev. B
\textbf{61}, 6213 (2000).

\bibitem{deisenhofer} J. Deisenhofer, M. V. Eremin, D. V. Zakharov, V. A. Ivanshin,
R. M. Eremina, H.-A. Krug von Nidda, A. A. Mukhin, A. M.
Balbashov, and A. Loidl, cond-mat/0108515.

%\bibitem{hagedorn}  F. B. Hagedorn and E. M. Gyorgy, Phys. Rev. {\bf 174},
%540 (1968).

\bibitem{mukhin97} A. A. Mukhin, V. D. Travkin, S. P. Lebedev, A. S. Prokhorov, A. M. Balbashov, and I. Yu.
Parsegov, J. de Physique IV (Colloque) \textbf{7}, 713 (1997); A.
A. Mukhin, V. D. Travkin, S. P. Lebedev, A. S. Prokhorov and A. M.
Balbashov, J. Magn. Magn. Mater. \textbf{183}, 157 (1998).

\bibitem{magneto} R. L. Walker, Phs. Rev. \textbf{105}, 390
(1957); P. Fletcher, I. H. Solt, Jr., and R. Bell, Phys. Rev.
\textbf{114}, 739 (1959); R. W. Damon and J. R. Eshbach, J. Phys.
Chem. Sol. \textbf{19}, 308 (1961); for a review see P.
R\"{o}schmann and H. D\"{o}tsch, Phys. Stat. Sol. (b) \textbf{82},
11 (1977).

\bibitem{morrish} A. H. Morrish, The physical principles of
magnetism, (Wiley, New York, 1965), p.556.

\bibitem{wang} S. T. Wang and C. W. Searle, Can. J. Phys. \textbf{49}, 387
(1971); S. E. Lofland, V. Ray, P. H. Kim, S. M. Bhagat, M. A.
Manheimer, and  S. D. Tyagi, Phys. Rev. B \textbf{55}, 2749-2751
(1997); A. Schwartz, M. Scheffler, and S. M. Anlage, Phys. Rev. B
\textbf{61}, R870 (2000).

\bibitem{zhou01} J.-S. Zhou and J. B. Goodenough, Phys. Rev. B \textbf{64}, 024421
(2001).

\bibitem{doloc} L. Vasiliu-Doloc, J. W. Lynn, A. H. Moudden, A. M. de Leon-Guevara, and A.
Revcolevschi, Phys. Rev. B \textbf{58}, 14913 (1998).

\bibitem{nagaev79} E. L. Nagaev, \textit{Physics of Magnetic Semiconductors} (Moscow, Mir Publ.,
1979); M. Yu. Kagan, D. I. Khomskii, M. V. Mostovoy, Eur. Phys. J.
B \textbf{12}, 217 (1999).

\end{thebibliography}
\end{document}